\documentclass{aa}
\usepackage{graphicx}
\begin{document}
\title{Astrometry with {\it Carte du Ciel} plates, San Fernando zone. \\
I. Digitization and measurement using a flatbed scanner.}

\author{B. Vicente \inst{1,2},
C. Abad \inst{2},
\and F. Garz\'on \inst{1,3}}

\offprints{B. Vicente}

\institute{Instituto de Astrof\'{\i}sica de Canarias (IAC),
La Laguna (S/C de Tenerife), E-38200 Spain.\\
\email{bvicente@iac.es}
\and
Centro de Investigaciones de Astronom\'{\i}a (CIDA),
Apdo. 264, M\'erida, 3101-A Venezuela.\\
\and
Departamento de Astrof\'{\i}sica, Universidad de La Laguna,
La Laguna (S/C de Tenerife), E-38200 Spain.
}

\date{Received 30 November 2006 / Accepted 16 May 2007}


\abstract
{The historic plates of the \textit{Carte du Ciel}, an
international cooperative project launched in 1887, offers
valuable first-epoch material for determining of absolute proper
motions.}
{We present an original method of digitizing and astrometrically
reducing \textit{Carte du Ciel} plate material using an
inexpensive flatbed scanner, to demonstrate that for this material
there is an alternative to more specialized measuring machines
that are very few in number and thus not readily available. The
sample of plates chosen to develop this method are original
\textit{Carte du Ciel} plates of the San Fernando zone,
photographic material with a mean epoch 1903.6, and a limiting
photographic magnitude $\sim$14.5, covering the declination range
of $-10^{\circ} \leq \delta \leq -2^{\circ}$.}
{Digitization has been made using a commercial flatbed scanner,
demonstrating the internal precision that can be attained with
such a device. A variety of post-scan corrections are shown to be
necessary. In particular, the large distortion introduced by the
non-uniform action of the scanner is modelled using multiple scans
of each plate. We also tackle the specific problems associated
with the triple-exposure images on some plates and the
\textit{r\'eseau} grid lines present on all. The final measures
are reduced to celestial coordinates using the Tycho-2 Catalogue.
}
{The internal precision obtained over a single plate, $3\mu m \sim
0\farcs18$ in each axis, is comparable to what is realized with
similar plate material using slower, less affordable, and less
widely available conventional measuring machines, such as a PDS
microdensitometer. The accuracy attained over large multi-plate
areas, employing an overlapping plate technique, is estimated at
0$\farcs$2. }
{The techniques presented here for digitizing photographic
material provide a fast and readily available option for the
exploitation of old plate collections. Our demonstration area,
consisting of $\sim$560\,000 stars at an average epoch of 1901.4
is presented as a practical example of the developed scanning and
reduction methods. These results are currently being combined with
modern astrometry to produce an absolute proper-motion catalogue
whose construction is underway. }


\authorrunning{Vicente, B. et al.}
\titlerunning{Astrometry with CdC-SF plates using a flatbed scanner.}

\maketitle

%

\section{Introduction}

The {\it Carte du Ciel} project was established at the
Astrophysical Congress held in Paris in 1887 and had a twofold
objective: the construction of a complete catalogue to V$\sim$11,
the {\it Astrographic Catalogue}, and to map the sky to V$\sim$14,
the {\it Carte du Ciel}. A total of twenty observatories around
the world were assigned the task of taking the photographic
plates. It was agreed that all observatories should use the same
design and the scale was a roughly consistent one arcmin/mm. More
details about the international congress can be found in Gill
(1898). Some participants urged that all the measuring should be
concentrated in one central bureau, but it was eventually agreed
that the measurement and reduction of the plates should be left to
the individual observatories. This plate material constitutes the
first observational full-sky record, currently with 100 years of
antiquity in most cases. As such, it presents a valuable resource
for wide-area proper motion determinations and, thus, kinematic
studies of the Galaxy. The {\it Astrographic Catalogue} objective
was successfully completed, culminating in the recent AC2000
Catalogue (Urban et al. 1998) on the Hipparcos reference system.
However, only a few observatories completed their assigned
declination zones for the {\it Carte du Ciel} project.

Some earlier attempts at scanning and reducing of individual
plates have been made yielding internal accuracies ranging from
0$\farcs$1 to 0$\farcs$2 using conventional measuring machines
such as a microdensitometer (Geffert et al. 1996, Lattanzi et al.
1991). More recent studies have made use of a handful of plates to
determine proper motions for specific objects of astrophysical
importance (e.g. Dick et al. 1993, Ortiz-Gil et. al 1998).

Recently, Rapaport et al. (2006) have reported on the construction
of a catalogue based on 512 plates in the Bordeaux Carte du Ciel
region, using the APM Cambridge automatic measuring machine, with
an estimation of the measurement error of about 0$\farcs$15
 (2.5 $\mu m$).

Because of the limited availability of conventional astronomical
measuring machines and the costs involved in their use,
alternative digitization strategies are worth exploring. The use
of small machines for scanning permits researchers to overcome the
reluctance of proprietary institutions to lend their plate
material, which in the present case is honoured as `historic
property' and, as such, is subject to stringent access control.
Digitization can thus only proceed on site. Trials employing a
flatbed scanner have been made (Lamareille et al 2003), but solely
for the purpose of judging the photometric precision attainable,
not astrometry.

The high speed of a scanner is afforded by imaging with a
multi-element detector, such as a 1D CCD array. This requires that
a large area be illuminated and imaged simultaneously, leading to
significant scattered light and the unavoidable decrease in
signal-to-noise and dynamic range of the scanned image. The lower
signal-to-noise will adversely affect both the detection limit and
the astrometric precision of detected images.

Although scanners are capable of spatial resolutions comparable to
a PDS, they are inferior in terms of their stability and
repeatability. The primary difficulty in digitizing with a
commercial scanner is the large distortions that are introduced by
the mechanical limitations of the scanner itself. Scanners are not
designed with the high-precision tolerances associated with a more
appropriate special-purpose measuring machine such as a PDS. Thus,
a detailed analysis and evaluation is necessary to ascertain if
astrometrically useful precision can be achieved using a scanner
and, if so, what reduction procedures are required.

In this paper we report on the methods developed in the
digitization and astrometric calibration of 420 \textit{Carte du
Ciel} plates, which have a mean epoch of 1901.4. Eventually, the
resulting early-epoch positions will be used to derive absolute
proper motions to the magnitude limit of the CdC plates by
combinating them with modern positions from the UCAC2 Catalogue
(Zacharias et al.~2004). This paper will confine itself to a
description of the techniques developed to process the scanner
measurements and an evaluation of the precision attained. However,
estimates of the final astrometric precision achieved will be
presented in terms of their impact on the planned proper-motion
measures and on the projected final proper-motion errors, as this
is the primary scientific motivation for the overall project.

\section{Plate material}

For its part of the {\it Astrographic Catalogue/Carte du Ciel}
project (AC/CdC in what follows), the Observatorio de San Fernando
(C\'adiz, Spain) was assigned the area between $-2\ ^{\circ}$ and
$-10\ ^{\circ}$ declination. The area was fully completed for both
surveys producing a total of 2520 plates. It is of note that the
collection of 1260 {\it Carte du Ciel} plates has not been
exploited up to now. The present digitization program was carried
out to do this.

Plates of the San Fernando zone were taken between 1892 and 1930,
using the Gautier Astrograph with an approximate scale of 60$''\
mm^{-1}$. Each plate covers a field of $2^{\circ} \times
2^{\circ}$ and observations were planned in a full overlapping
strategy, such that every star would be included in two different
plates, one of which would contain three exposures. Plates along
odd declinations were exposed three times, each 30-minute exposure
being shifted $7''$, producing a pattern of images for each star
that is roughly an equilateral triangle. Plates along even
declinations contain a single exposure lasting 30 minutes. All
of the plates in the {\it Carte du Ciel} project also contain a
superimposed {\it r\'eseau} grid of 27 perpendicular horizontal
and vertical lines, with 5 mm separation. These were included as
an aid to assist in the process of visual measurement.

\begin{figure}[h!]
\centering
\includegraphics[angle=+90,width=9.0cm]{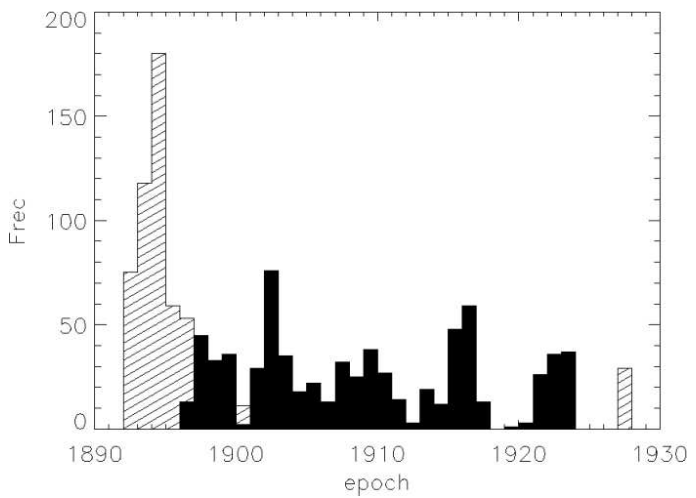}
\caption{Distribution of epochs of {\it Carte du Ciel} plates, San
Fernando zone. Bars with hatching represent even declination
plates, while black-filled bars correspond to odd declination
plates, i.e., triple-exposure plates.}
\label{fig01}
\end{figure}

\begin{figure*}[ht!]
\centering
\includegraphics[height=5cm]{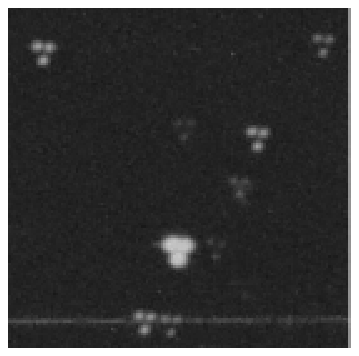}
\includegraphics[height=5cm]{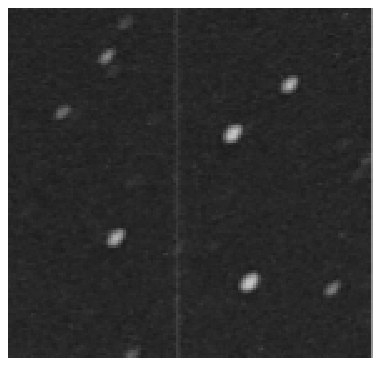}
\caption{\small Sample star images from a triple-exposure plate
(left) and from a single-exposure plate (right). These stars were
selected to sample different areas of their respective plates.
There is an obvious difference in image quality between the centre
(left) and the edge of a plate (right), where the effect of coma
becomes significant. R\'eseau grid lines can also be seen.}
\label{fig02}
\end{figure*}

Figure \ref{fig01} shows the distribution of epochs for the San
Fernando plates collection. Some examples of simple-exposure and
triple-exposure plates can be seen in Fig. \ref{fig02}.

\section{Digitization of the plates}

The merits of the photographic medium for astrometric work are
many, but among the drawbacks is the need to measure the positions
of images on the source material by additional effort. Nowadays,
digitization has become the only valid method of doing so, which
is important in view of the large amount of material comprising
the \textit{Carte du Ciel}. This digitization is typically done
with a specialized measuring machine, for example a PDS
microdensitometer. Such instruments are confined to a handful of
institutions, where they are permanently installed, i.e.,
immobile.

\begin{figure}[h!]
\includegraphics[width=8.1cm]{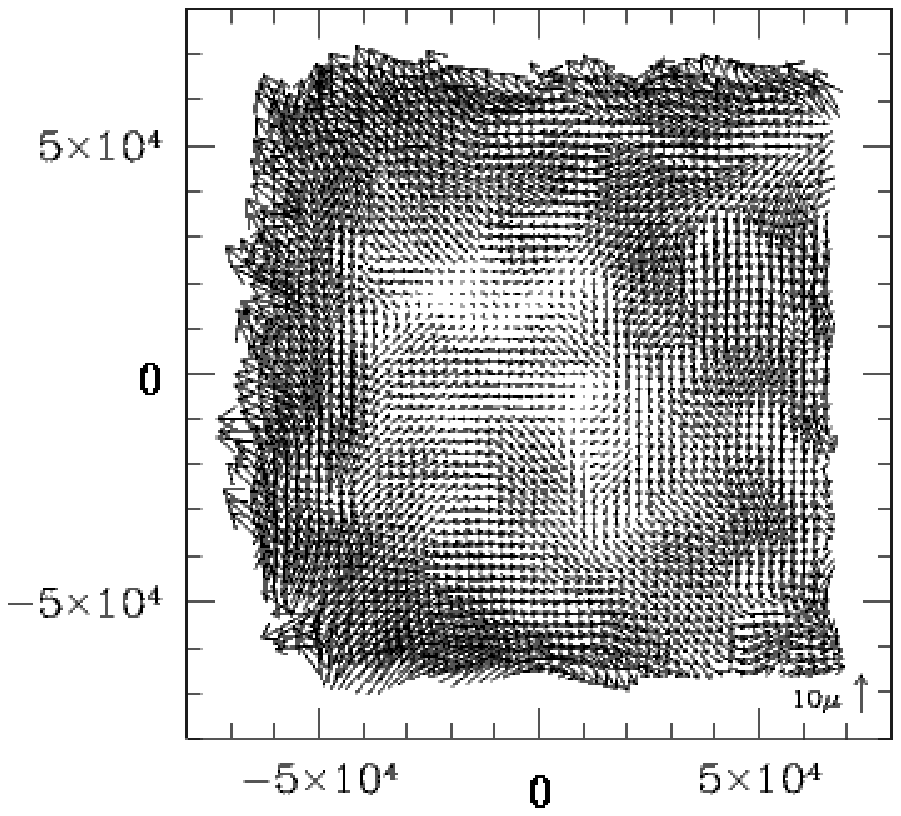}
\includegraphics[width=8.1cm]{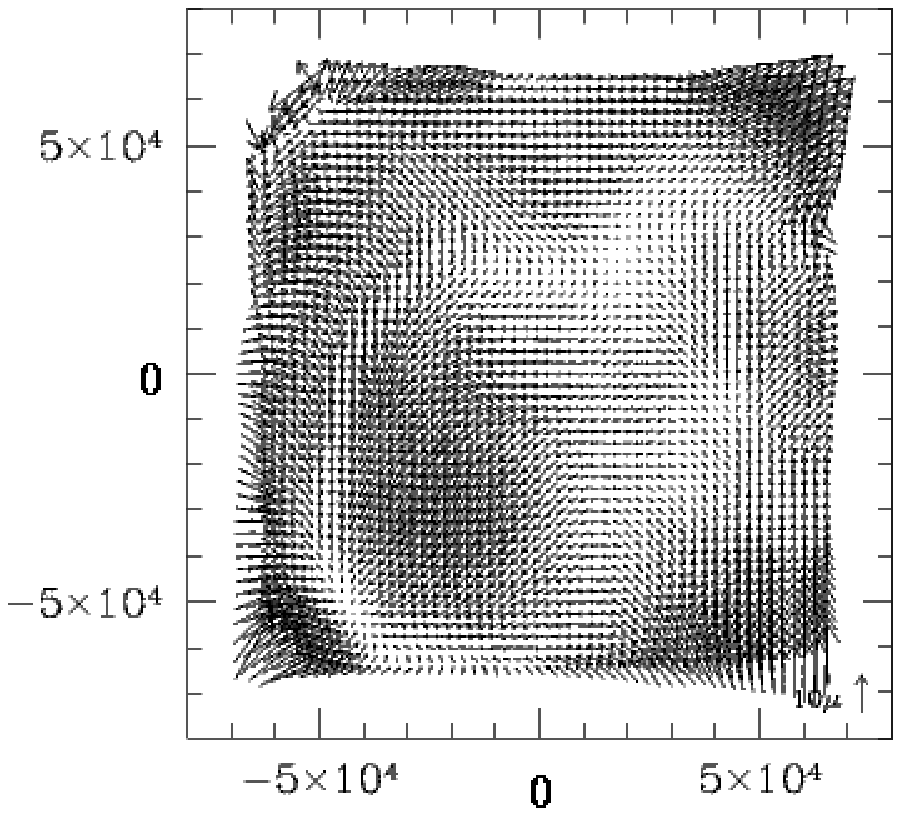}
\caption{The common distortion pattern found in the acetate copies
(top), and the residual distortion pattern remaining in an
individual acetate copy (bottom) after removal of the common
pattern. The scale of the residuals is indicated in the lower
right of each panel.}
\label{fig03}
\end{figure}

The photographic plates of the AC/CdC surveys represent an
important historic legacy of San Fernando Observatory and as such
cannot be removed from the observatory. Thus, it is not possible
to transport them to a PDS location for measuring. As an
alternative, the AC/CdC plates were duplicated onto acetate
substrate in 1999 in order to be measured with the PDS
microdensitometer of the Centro de Investigaciones de Astronom\'ia
(CIDA) in Venezuela. Preliminary tests involving the acetate
copies were performed, comparing measures from the PDS at CIDA
($\sim1.5\mu m$ internally) and those from the PDS at Yale
University (repeatability of $\sim0.6\mu m$). External comparison
to the original measures of the AC plates (described in Urban et
al. 1998) indicated that the duplication process and acetate
material introduced relatively large systematic errors, up to
$\sim15\mu m$ in amplitude. These inflated the overall standard
error of a single measurement to $\sim5\mu m$. Typical distortion
patterns are shown in Fig. \ref{fig03}.

In addition to the lower internal precision of the PDS at CIDA,
its inefficient serial scanning limited its production to just one
plate per day. Both factors, slow speed and errors introduced
during the duplication process, were incentives for investigating
the use of other digitization devices, such as flatbed scanners,
which are readily available and easy to transport. A portable
scanner has the advantage of allowing the measurement of the
original {\it Carte du Ciel} plates at San Fernando Observatory.
The high speed of the scanner allows repeated digitization, should
this prove of benefit in improving the final precision of the
measures. Preliminary studies were done with an Agfa DuoScan
scanner of the Universidad de Zaragoza (Spain) to study the
astrometric potential of a commercial scanner (see Vicente \& Abad
2003 for more details). We concluded that the combination of
flatbed scanner and original plate material yields astrometric
precision comparable to the PDS measurement of acetate copies, if
not better.

The scanner used in the current study is an Agfa DuoScan model
f40. It is a flatbed one-pass scanner with an optical resolution
1200 ppi $\times$ 2400 ppi, dynamic range of 3.0 in density, and
16 bits of digital resolution. It uses a trilinear CCD with
10\,600 elements. The important technical specification of this
particular model is its built-in scanning bed for transparencies.
The lower platform for transparencies has the advantage that
images scanned from it are captured directly, not through a glass
platter as is the case for opaque material.

We expect the scanner to introduce significant systematic errors,
which will differ in magnitude and degree of stability along the
two different axes. This expected difference is because of the
physical mechanism and manner in which the scanner operates. A
linear solid-state detector defines the $x$-axis of the system and
rides on a carriage that travels along the $y$-axis. By scanning
each plate in two orientations, rotated by $90^{\circ}$, we are
able to detect and separate the systematic errors introduced in
both axes' coordinates by the scanner. Details of the procedure
used are given in Sect. 5.

The Real Instituto y Observatorio de la Armada in San Fernando
(ROA, Spain) has completed the digitization of its collection of
2520 AC/CdC plates. The plates were scanned in 2003 in density
mode at the maximum resolution (10.5$\mu$m per pixel = 0$\farcs$63
per pixel), with scans of 13\,100$\times$13\,100 pixels
(2$\fdg3\times2\fdg3$), covering the desired area for each plate,
2$^{\circ}\times2^{\circ}$. Each plate was scanned twice and being
rotated 90$^{\circ}$ between scans. For each plate, the pair of
scans, which we refer to as scans A and B, are made immediately
one right after the other. A mechanical wooden holder was
manufactured to ensure that the same area of each plate would be
imaged during each of the two scans. The scan images were saved in
two-byte integer FITS format and recorded to CD-R disks.

\section{Measurement of stellar images}

The San Fernando collection of plates is complete and in
moderately good condition, considering the long time they have
been in storage. There are several characteristics of these plates
that complicate the determination of precise positions of their
stellar images: 1) the merging of the triple-exposure images on
the odd-numbered declination plates - especially difficult are
bright stars and those in the outer regions of the plate where the
coma is significant, 2) the blending and confusion of stars that
fall on r\'eseau grid lines, and 3) the false detections due to
plate flaws, spurious dust, and degradations that have accumulated
during storage. In addition to these, one must deal with the
typical problems of optical aberrations that are also present.

For the present study, one third of the full San Fernando CdC
collection has been analysed, covering the right ascension range
$\alpha=(06^h,14^h)$, comprising 420 plates (180 simple-exposures
plates and 240 triple-exposures plates).

\begin{figure}[h!]
\centering
\includegraphics[angle=0,scale=1.0]{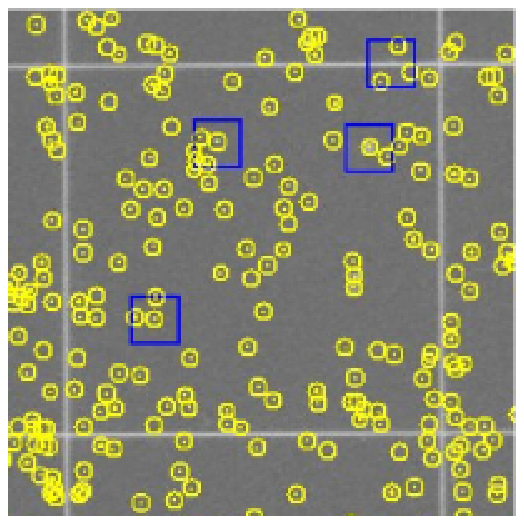}
\includegraphics[angle=-90,scale=1.0]{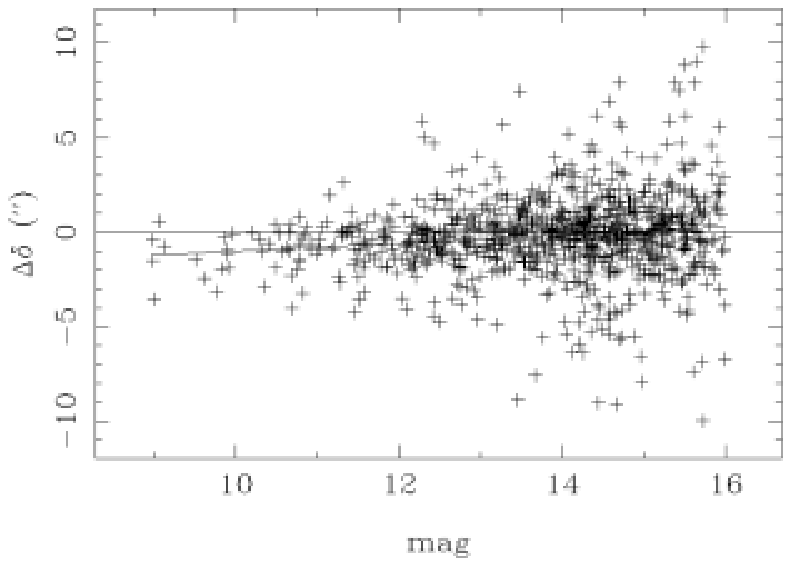}
\caption{Shortcomings of the raw SExtractor detections and
positions for the CdC plate material. In the upper panel, the
large number of spurious detections is illustrated for a
$5\times5\ mm^2$ section of a sample plate in which there are only
four real stars (squares). In total on this plate, 82\,000
detections were found, while only 1618 are actual stars. An
additional drawback of SExtractor-determined centres, as shown in
the lower panel, is magnitude equation in the positions as
demonstrated by the differences in positions calculated with
SExtractor and with Gaussian fitting. Only differences in
declination are displayed, the right ascension coordinate
exhibiting a similar behaviour.} \label{fig04}
\end{figure}

The reduction process begins with an initial detection and
centroiding of possible stellar images in each scan. This task is
accomplished using the software package SExtractor (Bertin \&
Arnouts 1996). The resulting list of detections includes a large
number of false detections, due to the numerous flaws, even after
those associated with the r\'eseau grid are removed. Figure
\ref{fig04}a illustrates the real-star detection efficiency for
the SExtractor threshold parameters adopted.

\begin{figure*}[ht!]
\centering
\includegraphics[angle=-90,width=16cm]{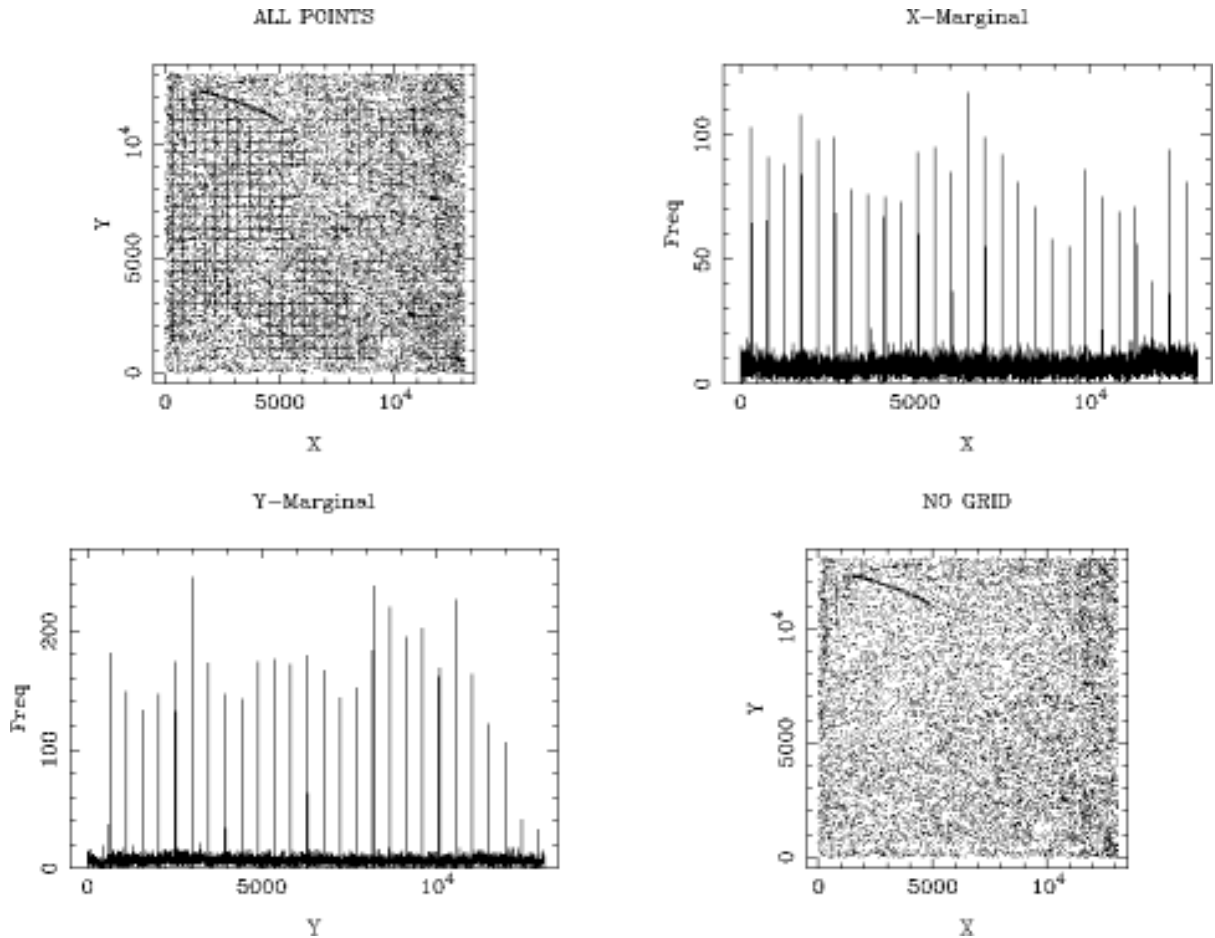}
\caption{ R\'eseau grid-line detection and removal for a sample plate.
Plotted are the $x$,$y$ positions of the SExtractor
detections before (upper left) and after (lower right) removal of
those that our procedure identifies with the grid. The procedure
makes use of the $x$- (upper right) and $y$- (lower left) marginal
distributions to deduce the location of the grid lines.}
\label{fig05}
\end{figure*}

We note that the SExtractor determined positions are simple the
photographic-density-weighted centre of light of the image. This
of the image. This is not an optimal determination of the center
of a stellar image, both in terms of random and systematic
measuring errors. Auer \& van Altena (1978) studied stellar image
centering in detail with photographic material. They found that 2D
Gaussian fits provided the most precise centres, being slightly
better than 1D Gaussian fits of the marginal distributions.  They
also established that centroids, moments of the distribution, are
less precise than functional fits to the stellar profiles. In
their case study, centroids provided a precision of
$(\sigma_{x},\sigma_{y}) = (1.8\mu m,1.6\mu m)$ compared to
$(\sigma_{x},\sigma_{y}) = (1.4\mu m,1.3\mu m)$ using
Gaussian-fitting. In addition, for bright stars the nonlinear
photographic register combines with asymmetric image profiles (due
to aberration and/or guiding error) and leads to magnitude
equation. Magnitude equation is the undesired correlation between
the position of the image centre and the magnitude of the star
producing the image. It is caused by the combination of an
asymmetric image profile and the non-linear response of the
photographic detector.

SExtractor is well-suited to our large fits images in that it is
quite fast and consistently provides reliable image detections.
The SExtractor centroids are an appropriate choice for the task of
mapping the r\'eseau grid lines and providing an approximate
astrometric solution to each plate, but not as the basis for our
ultimate astrometry. Therefore, we choose to refine the centres
using a bi-variant Gaussian fitting method, developed at Yale for
use with their PDS machine (Lee \& van Altena 1983). Some
comparative tests show evidence of the improvement derived from
the Gaussian-fitting versus SExtractor centroiding. A magnitude
equation is found in the differences of positions between
Gaussian-fitting and SExtractor centroids (Fig. \ref{fig04}b). We
have investigated the source of this trend and find that it
appears in differences of the SExtractor positions compared with
an external catalogue, but not in differences of the Gaussian-fit
positions with this catalogue. This reinforces the conventional
notion that simple centroids are not the optimal centering method
for photographic material. Even the Gaussian centering algorithm
does not remove all the systematic errors as a function of
magnitude, so we will still have to study the magnitude effect in
the reduction process, as will be shown in Figs. \ref{fig16} and
\ref{fig17}.

The Yale Gaussian-fitting code requires an initial input position
for each star. We used an external star catalogue to provide these
input positions, thus addressing several issues; primary among
these is that of cleaning, by only attempting to centre objects
known to be stars. The SExtractor positions help in defining the
projection of the catalogue, at the epoch of the plate material,
onto the $x$,$y$-system of each plate so the Gaussian centering
can be performed at the projected star locations.

The input catalogue we use is the UCAC2 -- The Second USNO CCD
Astrograph Catalogue (Zacharias et al. 2004). UCAC2 is the logical
choice as it will also be used to provide the second-epoch
positions that will eventually be combined with our CdC measures
to calculate proper motions.  At present, it represents the most
precise astrometry ($\sim$0\farcs020 -- 0\farcs075) available that
reaches to the magnitude limit of the CdC plate material.

In practice, UCAC2 stars are identified within the list of
SExtractor detections by positional coincidence, and their
SExtractor centroids are used to determine a 4$^{th}$-order
polynomial plate solution by least-square fitting. The plate
solution then allows the full list of UCAC2 celestial coordinates
to be properly projected onto the $x$,$y$-system of the plate
scan. These projected $x$,$y$'s are then used as approximate input
positions to determine refined centres for all UCAC2 stars that
appear in the plate scan. This process of Gaussian centering with
an input list from UCAC2 positions is done for both simple and
triple exposure plates.

\subsection{Elimination of grid lines}

The presence of the {\it r\'eseau} on the plates creates a large
number of non-stellar detections by SExtractor along the grid
lines. These can be identified geometrically and then eliminated
from the list of detections so as not to confuse and spoil the
plate solution and the subsequent UCAC2 projection onto the plate.

Therefore, a method for eliminating the grid lines has been
developed. The marginal distribution along the $x$ scan axis is
formed by calculating the binned distribution of $x$ coordinates
of all detections on the plates. Similarly, a $y$ marginal
distribution is formed. Sample marginal distributions are shown in
Fig. \ref{fig05}.

The rectangular grid is sufficiently well aligned to the $x$ and
$y$ axes of the scan, such that a pattern of peaks is seen in the
marginal distributions caused by the large number of detections
along the grid lines. Knowing the nominal spacing of the grid
lines (roughly 480 pixels), it is relatively straightforward to
detect their presence in the marginal distributions in an
automatic way. Basically, one starts with the highest peak in the
distribution, then searches for other peaks around the known
distance from the previous peak. Once the $x$ and $y$ locations of
the grid lines are determined, all points within $\pm$1 pixel of
the grid lines are assumed to be a spurious grid detection and are
flagged as such.

From a visual inspection of the plates, it is seen that the
scanning orientation is not, in general, perfectly aligned with
the grid lines on the plate.  They differ by a small rotation
angle. The code calculates and includes this angle in its mapping
of the grid lines, allowing the grid detections to be eliminated
without the need to rotate the original $(x,y)$ positions.

\subsection{Treatment of triple-exposures plates}

On the plates with triple exposures, the three images of a bright
star will blend to form a central blob that is detected as a
single image by SExtractor. The automatic matching between UCAC2
and SExtractor positions is heavily weighted by the brightest
stars in both lists, so that the plate solution it produces
provides the transformation between UCAC2 coordinates and the
centre-of-light of the three exposures. In order to predict
locations on the plate for the other three exposure systems, the
relative positions of these three images with respect to the
centre-of-light position must be determined.

\begin{figure}[h!]
\centering
\includegraphics[angle=+90,width=4.3cm]{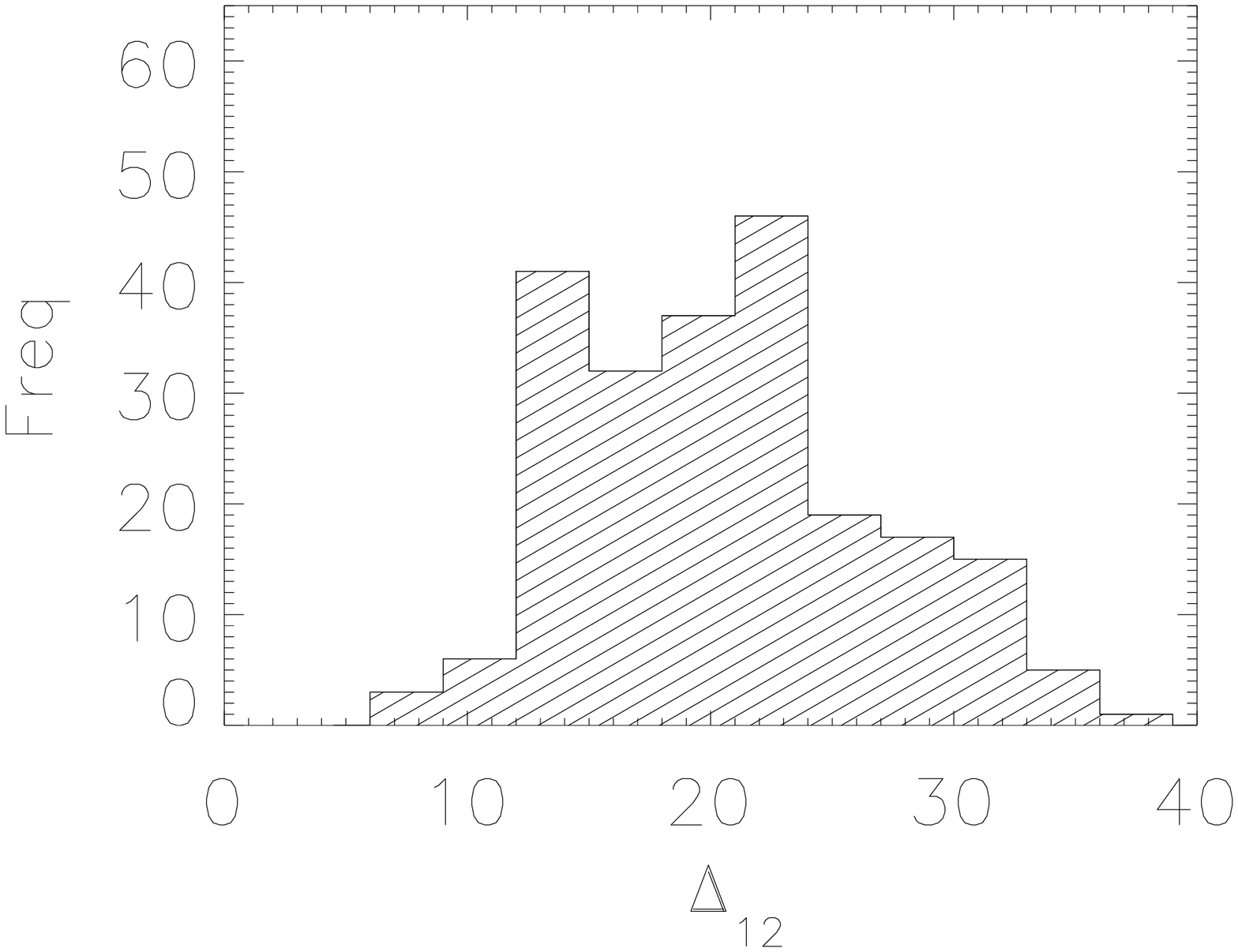}%
\includegraphics[angle=+90,width=4.3cm]{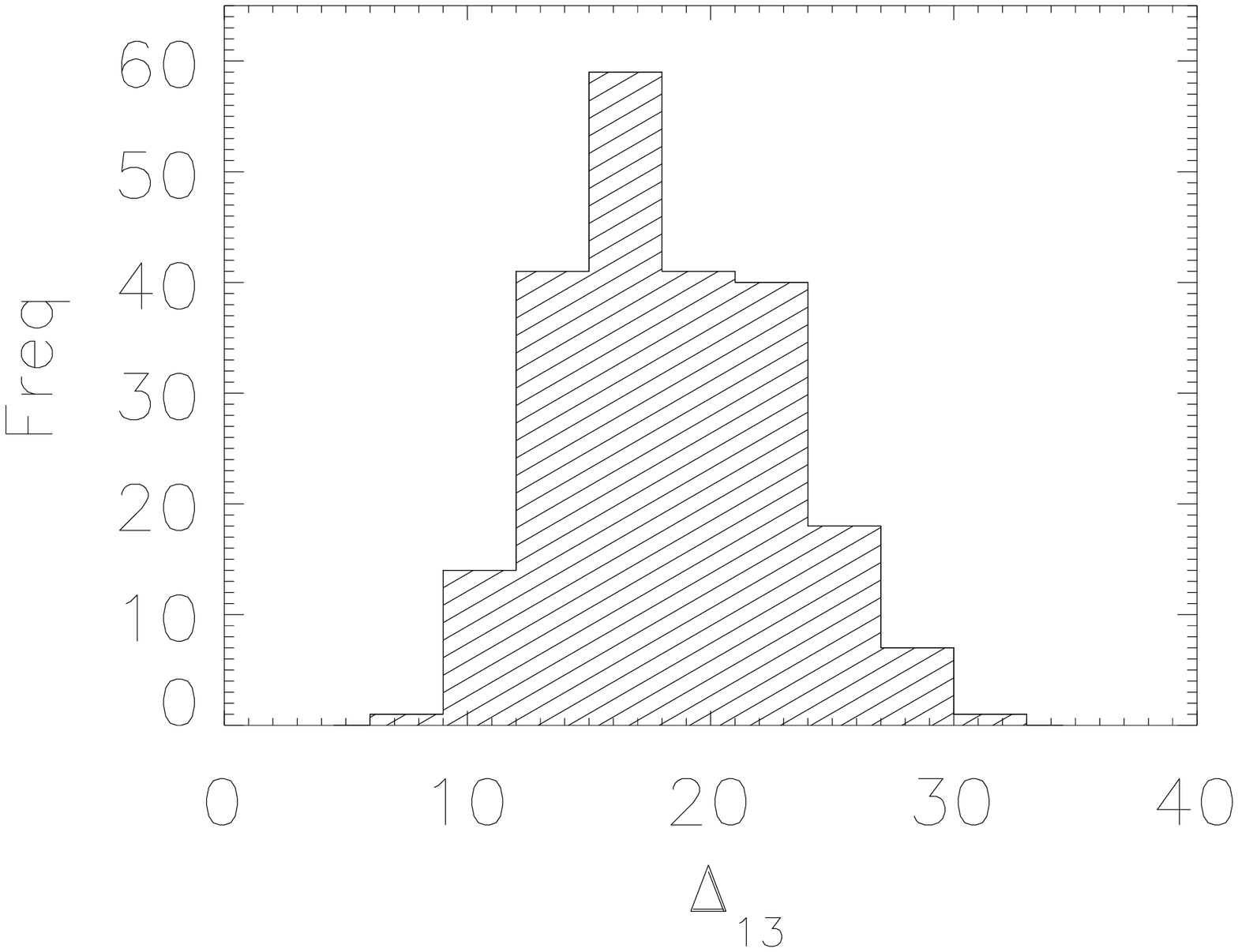}
\caption{Distribution of relative separations in pixels between
exposures 1 and 2 ({\it left}) and between exposures 1 and 3 ({\it right}) for
223 triple-exposures plates. We can see that not all the plates
have equally spaced offsets so it is necessary to calculate them
individually for each plate.}
\label{fig06}
\end{figure}

The telescope offsets used during the observations attempted to
place the three images in a roughly equilateral triangular
pattern, in theory $7''$ on a side, or 12 pixels in our scans.
However, the offsets vary significantly (Fig. \ref{fig06}) from
plate to plate and must be determined individually. There is also
an expected variation in the triangular pattern as a function of
position across each plate, due to the slightly different
telescope tangent point of each exposure.  This quadratic
(plate-tilt) variation is small enough that for the purposes of
identifying images by positional coincidence can be ignored. We
have developed a method for calculating the offsets of the three
another and to the blended centre-of-light system. It involves
calculating the $x$ and $y$ separations between each detection and
the other detections in its neighbourhood, and then searching for
high-density clustering within this 2D separation space. The
relative offsets of the three exposures, i.e., the vertices of the
triad exposures, are then deduced from the relative positions of
these clusterings.

For each plate we determine the relative offsets, $\Delta x_{ij} =
x_j-x_i$ and $\Delta y_{ij} = y_j-y_i$, where $i$ and $j$ range
from 1 to 3, the three vertices of the triple exposures.
Distributions of the exposure separations are shown in Fig.
\ref{fig06} where $\Delta_{ij} = \sqrt {\Delta x_{ij}^2 + \Delta
y_{ij}^2}$. Running the procedure on 223 triple-exposure plates,
the mean relative distances of the vertices are found to be
$<\Delta_{12}> = 20.6 \pm 6.2$ pixels and $<\Delta_{13}> = 18.3
\pm 4.6$ pixels. The large dispersions about the mean values
clearly show the need for separate offset calculations for each
individual plate.

\begin{figure}[h!]
\centering
\includegraphics[angle=90,width=7cm]{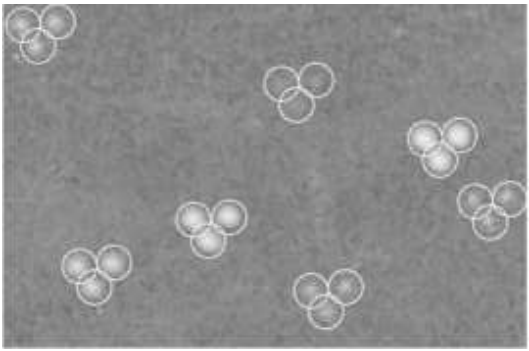}
\caption{
Predicted locations of UCAC2 star images, shown as circles,
on a sample area with four sets of measurable triple-exposure images.}
\label{fig07}
\end{figure}

These offsets are applied, in turn, to the predicted
centre-of-light locations of the UCAC2 stars in order to get the
location of these stars in each of the three offset exposure
systems (Fig. \ref{fig07}). The 2D Gaussian centering is then
performed using each of the three input lists. Because of
(variable) blending across the plates, not all the objects in the
input list center successfully. This will affect the completeness
of the final catalogue. A loss of up to 15\% of stars can be
expected due to interferences with the grid lines and spurious
flaws, as well as the blending of the triple exposures.

\section{Distortion introduced by the scanner}

As noted earlier, the imperfect nature of the scanner will deform
the scanned image, imparting a distortion into its measured
positions. As an aid in understanding the expected functional
dependencies, we present a heuristic model of the distortions
introduced, and they affect the true positions $(x_t,y_t)$, giving
rise to the measured positions $(x_{m},y_{m})$.

\begin{figure}[h!]
\centering
\includegraphics[width=7cm]{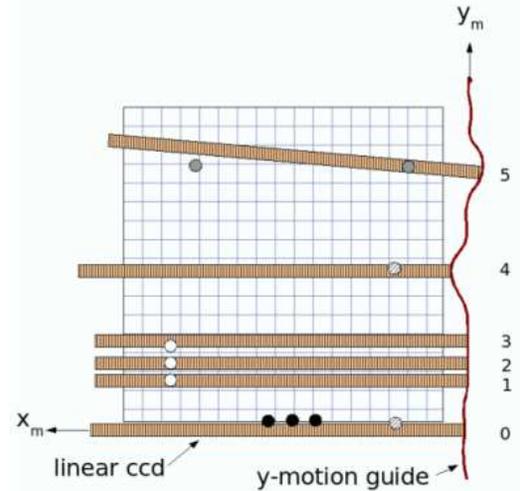}
\caption{ Schematic illustrating the operation of a flatbed
scanner with exaggerated flaws in its structure and motion, to
demonstrate the variety of distortions that are introduced, in
theory, by the scanning process. Different types of errors, with
their expected functional dependences, are shown at the various
labeled positions and explained in the text.} \label{fig08}
\end{figure}

Figure \ref{fig08} illustrates several of the expected distinct
types of scanner distortions.  Each type of deviation, labeled by
position in the figure, will have a functional dependence
associated with it as described in detail below.

\begin{itemize}
\item The sample stars shown in black are evenly spaced in
$x_{t}$, but unless the linear ccd is perfectly straight and has
ideal ruling, the measured $x_{m}$ values will not be evenly
spaced (see Position 0). This is an error in $x_{m}$ as a function
of $x$ (which may be equally well expressed as a function of
either $x_t$ or $x_m$).

\item The three white stars are evenly spaced in $y_{t}$, but the
motion of the ccd along $y$ is irregular. Although the scanner is
designed to move regularly from position 1 to 2 to 3, in fact it
skips and slips and measures the third star incorrectly (see
Positions 1, 2, 3). This type of displacement produces the most
significant distortion seen in our scanner, resulting in
large-amplitude errors in $y_{m}$ as a function of $y$.

\item The two stars marked with hatching have the same $x_{t}$
position, but because of the bump in the rail guiding the y-motion
of the ccd, when the second star is measured, its $x_{m}$ value
will appear lower. This is an error in $x_{m}$ that is a function
of $y$ (see Position 4).

\item Finally, the ccd might not remain exactly perpendicular as
it moves in $y$. The two grey stars have the same $y_t$
position, but the scanner will measure different values of $y_{m}$
because the stars are separated in $x$. This is an error in
$y_{m}$ that is function of $x$ (see Position 5).
\end{itemize}

The assumed stability of the solid-state detector, oriented along
the $x$-axis, suggests that the deviations described above are
separable along the two scan axes. Furthermore, by scanning each
plate in two orientations, rotated by 90$^{\circ}$, the large $y$
deviations can be corrected by comparison to the $x$ positions of
the complementary scan. The details of the correction procedure we
have developed are given below. The explicit use of subscripts t
and m meant here to distinguish between measured and true
coordinates. We will drop the m subscript at this point, since all
subsequent references to coordinates will be measurements.

\begin{figure*}[ht!]
\centering
\includegraphics[angle=-90,width=14cm]{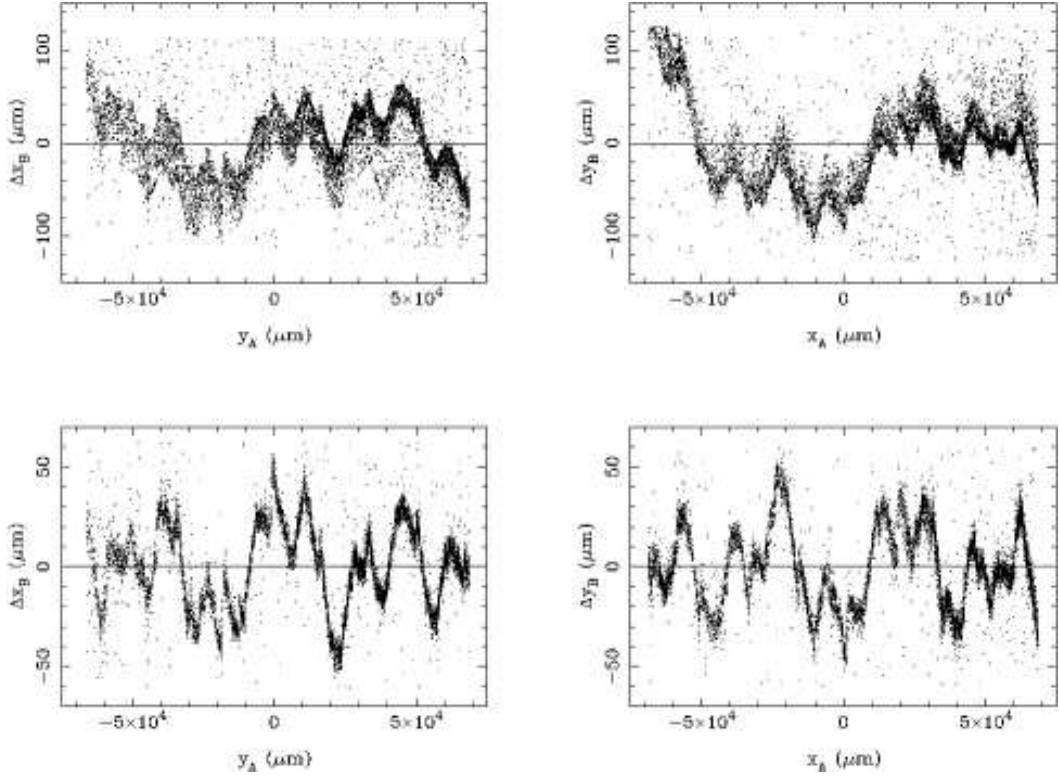}
\caption{ Sample comparison of differences in image positions
derived from two scans of the same CdC plate, rotated $90^{\circ}$
relative to one another. A linear transformation is performed to
align the two scans (upper panels) and residuals plotted versus
plate coordinates, as labeled. $\Delta x_B$ represents deviations
along the $x$-axis of scan B. These, in general, are derived from
differences between $x_B$ and transformed $y_A$ measures (and
similarly for $\Delta y_B$). Note that $x_A$ and $y_B$ are
anti-parallel measures along the same axis of the physical plate,
and similarly $y_A$ and $x_B$ are parallel. The large-amplitude
errors in the scanner's $y$-axis are readily visible. The 1D
function characterising the scanner $y$-axis measuring error is
seen more clearly in the lower panels after the removal of the low
frequency distortion in the upper panels, fitted as a cubic 2D
function.} \label{fig09}
\end{figure*}

\subsection{Scanner-distortion correction}

As discussed, the flatbed scanner is expected to introduce
substantial distortions into the scanned image, hence positions,
and these must be corrected before being transformed into
celestial coordinates. Based on the manner in which the scanner is
built and operates, we expect any image distortion along the
$x$-axis to be constant from scan to scan. That is, the systematic
error in $x$ as a function of $x$ will always be the same,
although we don't know the form of that distortion.  The
distortion in $y$ is expected to change with each scan, because of
unpredictable slippage as the carriage moves. We note that one
might expect some portion of the $y$-distortion to be stable from
scan to scan and, in fact, this is seen.  Yet there is still a
significant component that varies from plate to plate on top of
the common component, resulting in the total $y$ deviation.

This behaviour suggests a two-step procedure for removing the
distortion caused by the scanner. First, the presumed constant
metric of the scanner's $x$-axis will be utilised to remove the
deviations in both $y$ and $x$ caused by the unpredictable
$y$-motion of the scanner, by comparing rotated scans of the same
plate. This allows us to put both $x$ and $y$ measures of both
scans onto the metric defined by the $x$-axis of the solid-state
detector. Second, the remaining unknown but constant form of the
$x$-axis distortions, which by then will be present in both axes,
will be determined by comparison to an external source - the
independent $x,y$ measures from the contemporaneous AC plate
material described by Urban et al.~(1998). The constancy of the
$x$-distortion allows the residuals from the AC comparison to be
stacked for many fields, yielding a well-determined 2D final
correction mask that is applied to all plates.

An overview of our scanner-correction pipeline is as follows,
with details of the procedures described in the next two subsections:

\begin{itemize}
\item {correction of a deviation in $y$ as a function of $x$ by
comparing rotated scans of the same plate, implicitly adopting the
scanner's $x$-axis to define a stable system for both axes;}

\item {correction of a remaining deviation of $x$ as a function of
$y$, again using comparison of rotated scans;}

\item {correction of a roughly cubic polynomial distortion that is
common to all scans; followed by}

\item {correction of the deviations in the adopted $x$-axis system
from geometric linearity by comparison to independent, external AC
measures.}

\end{itemize}

\subsubsection{Internal scanner-distortion correction}

Utilising the presumed stability of the $x$-axis, we use the
repeated, 90$^{\circ}$-rotated scans A and B, to transform all $y$
measures onto the ``system'' of the scanner's $x$ measures, i.e.,
the metric defined by the solid-state detector. Residuals from a
polynomial transformation between scans A and B, shown in Fig.
\ref{fig09}, clearly show the 1D function that dominates the $y$
measure distortion for a typical plate. The function is
well-defined in the bottom panels that correspond to a general
cubic solution between booth scans. More precisely, these
residuals indicate that portion of the shape of the $y$ distortion
that is of a higher order than what can be described by the cubic
polynomial transformation model. The large cubic-polynomial
component turns out to be common among all scans, and thus we have
chosen to calculate and correct for it after we make the
corrections of the scan-to-scan varying deviations such as shown
in Fig. \ref{fig09}.  This common component to the distortions
will be tackled after the scan-to-scan variations are addressed.

The errors incurred by the motion of the carriage along $y$ are
erratic and range up to $50\mu$m in size, indicating that the
scanner has a significant problem with slippage and non-uniform
motion of the carriage. Our overall strategy for this
internal-correction step will be to first treat these
large-amplitude distortions and then to address successively
smaller ones, as they reveal themselves.

For a given pair of scans, there are two 1D $y$-axis functions to
be determined. These functions are essentially the differences
between scan A's $x$-axis and scan B's distorted $y$-axis and
between scan A's distorted $y$-axis and scan B's $x$-axis. The
appropriate residuals are used to define the 1-d functions of the
distortion pattern, employing the technique of ``Weighted Sliding
Polynomial" (Stock \& Abad 1988) to parametrize the functions.

\begin{figure}[h!]
\centering
\includegraphics[width=8cm]{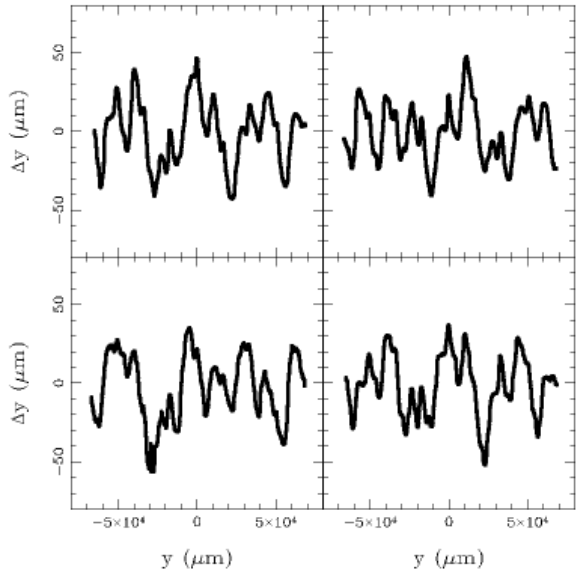}
\caption{Variation in the dominant $y$-axis scanner distortion over
time. While the $y$-axis deviation changes very little between
consecutive scans, over time the shape changes significantly. The
$y$-residuals between rotated scan pairs are shown for four
different plates, well separated in the time at which they were
scanned.}
\label{fig10}
\end{figure}

There is a high degree of similarity in the $y$ distortion pattern
of the successive scans A and B.
Over time, this $y$-distortion pattern changes substantially.
Figure \ref{fig10} displays the pattern for four different plates
scanned days to weeks apart.
It is for this reason that the $y$-distortion function for each scan
is derived separately, despite a superficial similarity between successive
scans.

\begin{figure}[h!]
\centering
\includegraphics[angle=-90,width=7cm]{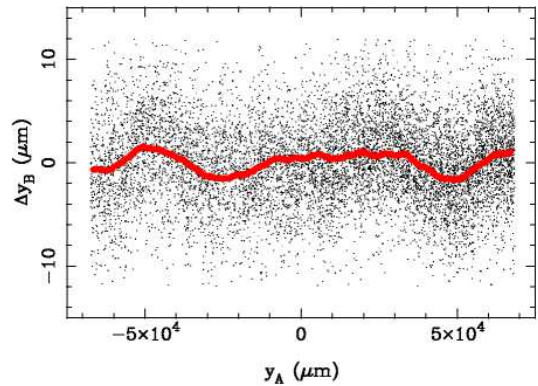}
\caption{
Detection and fitting of the smaller cross-axis deviations,
i.e., $x$-deviations as a function of $y$ coordinate,
after removal of the dominant $y$-axis distortion as a function of $y$.
}
\label{fig11}
\end{figure}

Once the 1D $y$-axis function is determined for each scan, it is
applied to the $y$ measures of the scan. (The underlying, common
cubic-polynomial component will be applied later.) Residual plots
from another cubic solution made subsequent to the $y$-axis
correction reveals a smaller, but easily measured, cross-axis
distortion, a deviation in $x$ as a function of $y$. An example of
this smaller deviation is shown in Fig. \ref{fig11}.

After application of this second 1D correction function (Fig.
\ref{fig11}), the residuals show no remaining discernible
systematics. Specifically, the other possible cross-axis
distortion, illustrated at position 5 in Fig. \ref{fig08}, does
not seem to be exhibited by this scanner. Thus, no correction for
this type of distortion is made.

\begin{figure*}[]
\centering
\includegraphics[width=5.4cm]{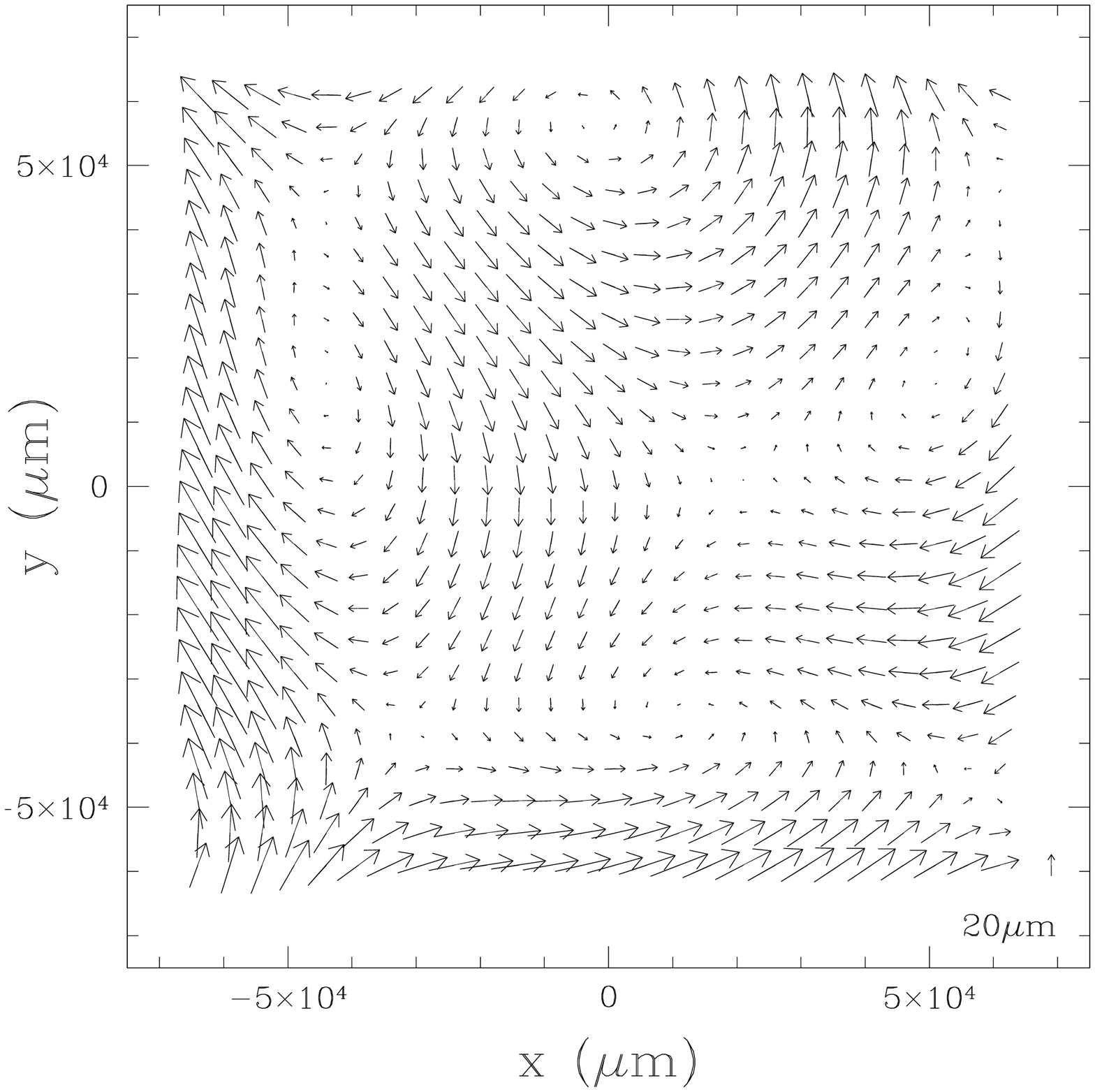}%
\hspace{0.5cm}
\includegraphics[width=5.4cm]{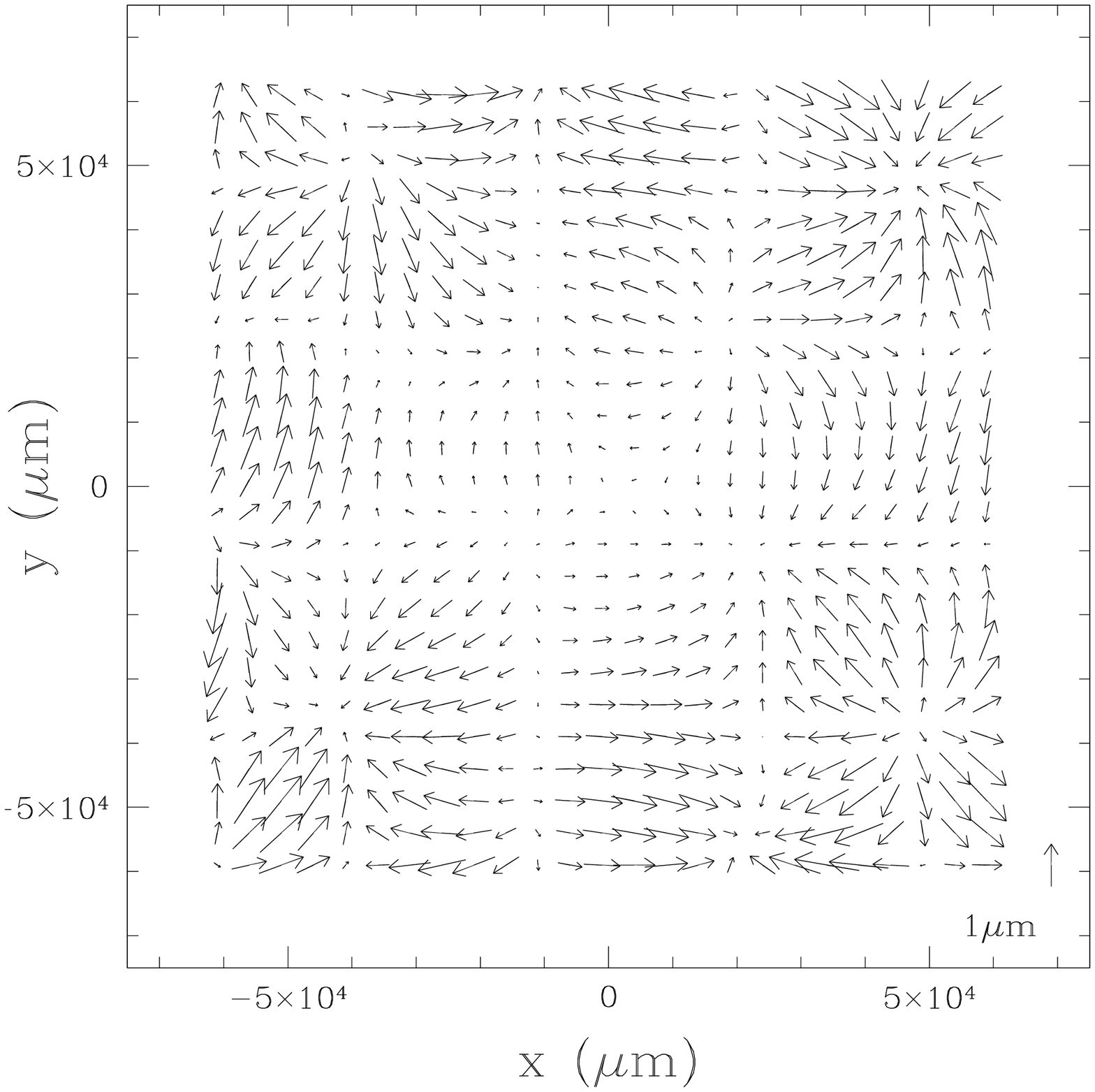}%
\hspace{0.5cm}
\includegraphics[width=5.4cm]{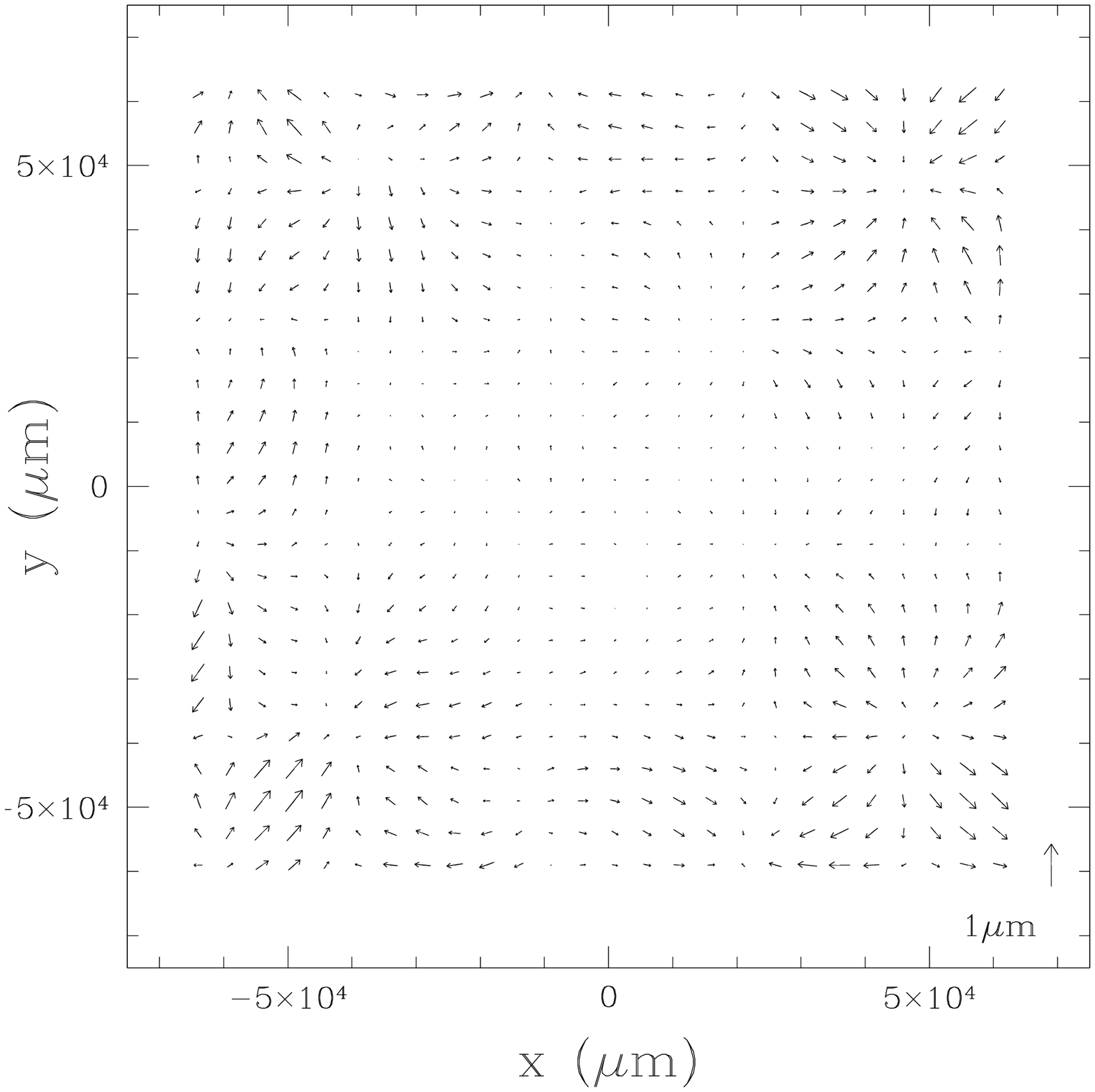}
\caption{\small Mean residual vectors as a function of position on
the plate before and after various internally-calculated
corrections are applied, comparing a sample plate's A and B scans
by linear transformation. At the left are the residuals based on
uncorrected scanner positions. In the middle are the residuals
that result from having corrected only for that portion of the
scanner distortion that is common between all scans and that can
be approximately represented by a general cubic function. To the
right is the residual plot after also subtracting the 1D
distortion functions specifically derived for the scans of this
plate.} \label{fig12}
\end{figure*}

Early tests indicated that the cubic-polynomial component of the
scanner distortion is stable and common to all scans. To be
precise, what we address here is the difference in a
cubic-polynomial deviation along the $y$-axis and along the
$x$-axis, since we only have our scan A/B pairs in these internal
comparisons to characterise it. Nonetheless, having recognised
this component as stable and common to all scans, it was decided
to calculate it based on stacked residuals from all scan A/B
pairs. We do so by first applying the two 1D (scan-to-scan
varying) corrections described above and then performing a simple
linear transformation between the resulting scan A and scan B
positions. The residuals from these linear transformations are
then stacked into a common 2D vector plot and a correction mask is
determined. This single mask, which effectively describes the
cubic-polynomial modelled in the previous steps, is then applied
to both scans of each plate. In practice, a second series of
linear solutions, residual stacking, mask construction, and
application is performed, to ensure that the systematics are
removed to the greatest possible extent.

To provide an illustration of the relative amplitudes of the
common (cubic) scanner distortion and the plate-to-plate-varying
1D deviation functions that reside on top of it, as well as to
demonstrate the validity of using a single mask to describe the
common component, Fig. \ref{fig12} shows the vector residuals for
a sample plate's A/B scan pair before and after each of these two
corrections have been made. The left panel indicates the residuals
from a linear transformation between scans A and B before any
corrections have been applied. The middle panel shows the
residuals if only the common component is corrected, as
represented by the mask constructed from the stacked residuals.
Note the factor of 20 change in plotting scale of the vectors. The
third panel shows the residuals after also correcting both scans
with their appropriate 1D functions along both axes. Figure
\ref{fig12} is for illustration purposes, as in practice the
corrections are not performed in this order. However, the vast
improvement seen by applying of the common mask alone, as well as
the lack of systematics in the residuals of the final
internally-corrected positions, validate that this component is
well-described by the common mask.

\begin{figure}[h!]
\centering
\includegraphics[width=8.0cm]{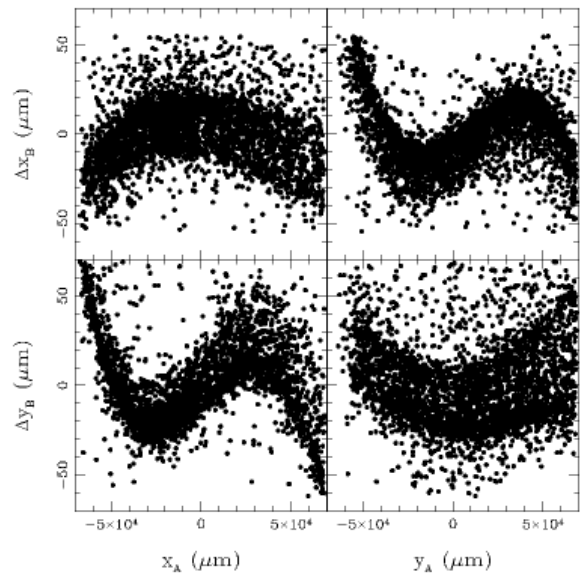}
\caption{ Residuals from a linear transformation between a scan-A
scan-B pair for a sample plate, after applying of the two 1D
correction functions. The remaining relative distortion between
the rotated scans is approximately cubic in form.  In practice, a
2D correction mask, constructed from stacked residuals of all
plates is used to correct for this remaining distortion.}
\label{fig13}
\end{figure}

Figure \ref{fig13} shows the magnitude of the common,
approximately cubic distortion in an absolute sense. Compare the
amplitude of the variations with those of the 1D deviations
illustrated in Fig. \ref{fig09}. If not for the inclusion of
general cubic terms when generating the residuals shown in Fig.
\ref{fig09}, the 1D deviation functions could not have been so
well-determined.

Summarising the internal portion of our scanner-distortion
correction procedure, Table \ref{table:1} shows the improvement in
precision attained by tracking the dispersion of positional
differences between rotated scan pairs at each step in the
process.

\begin{table}[h!]
\caption{RMS after each step in the scanner-distortion
correction.}
\label{table:1}
\begin{tabular}{l c c}
\hline
Step & RMS x ($\mu m$)& RMS y ($\mu m$) \\
\hline
 Initial (uncorrected) & 31.2 & 32.9 \\
 1-d function corrected  & 18.2 & 23.0 \\
 Correction cubic pattern & 4.61 & 4.42 \\
 2nd iteration cubic pattern & 4.52 & 4.31 \\
\hline
\end{tabular}
\end{table}

\subsection{External scanner-distortion correction}

At this point in the reduction pipeline, the $x,y$ measures will
have been corrected as much as possible using internal
comparisons. These internally-corrected measures will still
contain a systematic distortion that corresponds to the unknown
figure of the scanner's solid-state detector that defines the $x$
scan axis. Presumably, this distortion is a 1D function of
position, to which both axes have now been transformed and which
should be consistent from scan to scan. This remaining distortion
can only be determined by comparison to an external catalogue or
set of measures.

As external reference we use the existing $x,y$ measures of the
AC, described by Urban et al.~(1998) and used in the construction of
the AC2000 Catalogue.
These manual measures are of similar plate material, having been
taken with the same telescope as our CdC plates.
The original AC measures, in digital form,
were made available to us by the staff of the USNO.
Any systematic differences between the AC measures and
our internally-corrected positions will be dominated by the expected
remaining scanner distortion, i.e., that due to the geometric non-linearity
of the scanner's solid-state detector.
The AC measures are used to correct this distortion in the following
manner.

\begin{figure}[h!]
\centering
\includegraphics[width=7.5cm]{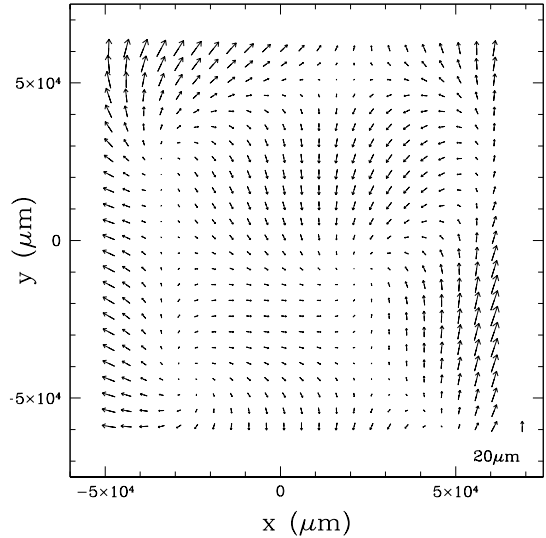}
\caption{\small
Stacked vector differences between internally corrected scanner positions and
the independent external set of measures for AC stars (Urban et al. 1998),
before correction of the scanner measures by a common-pattern mask.
The differences shown represent the sum of 340 scans (170 simple-exposure plates).
}
\label{fig14}
\end{figure}

Our internally-corrected scanner measures are transformed to the
system of the AC measures by a linear transformation. Initially,
the rms of the resulting residuals is $(rms_x,rms_y) = (11.3\mu
m,12.7\mu m)$. The residuals, as a function of $(x,y)$, is then
used to create a correction mask.  However, there are not enough
AC stars on any given plate to do so reliably. Therefore,
residuals from a large number of plates are stacked, a correction
mask is constructed from these, and then it is applied to every
plate's measures, (see Fig. \ref{fig14}). The resulting rms, after
correction for this common pattern, becomes $(rms_x,rms_y) =
(6.5\mu m, 7.1\mu m)$.

The estimated positional errors of the AC measures are a function
of declination zone (Urban et al.~1998). For the San Fernando
zone, the errors are about 0$\farcs$30 to 0$\farcs$35, which at
the scale of the plates (60$''\ mm^{-1}$) corresponds to 5 to 6 $\mu m$.
Thus, achieving an rms in the differences of $\sim 7\mu m$
indicates that our corrected scanner positions have errors
less than $\sim 5 \mu m$ for a single scan. In the following
section, a combined overlap solution will allow a better estimate
of the uncertainties in the corrected scanner positions.

Based solely on scan A/B residuals, though, the final resulting
single-measurement internal-error estimate for well-measured
stars, is

\begin{itemize}
\item
For simple-exposure plates the $x$ and $y$ uncertainties are
(3.18$\mu$m, 3.15$\mu$m) = (0$\farcs$19 , 0$\farcs$18).

\item
For the triple-image plates, the corresponding uncertainties are
(5.52$\mu$m , 5.09$\mu$m) = (0$\farcs$33 , 0$\farcs$31).
\end{itemize}

\section{Transformation to celestial coordinates}

Conversion to celestial coordinates and estimation of the
resulting precision, as well as the exploration of any remaining
systematic errors, are described in order to evaluate the
astrometric quality of the CdC-SF plate/scanner combination. The
scientific value of the final astrometry will be in its potential
usefulness in providing first-epoch positions for deriving proper
motions. Thus, we analyse the CdC-SF positions within this
context, while the construction of a proper-motion catalogue based
upon them is currently under construction.

A transformation from $(x,y)$ coordinates into celestial
coordinates ($\alpha,\delta$) has been made using the
block-adjustment technique (Stock 1981) including a determination
of the field distortion (Abad 1993). This technique utilises not
only external catalogue reference stars but also common images in
overlapping plates corresponding to the same star. A system of
link conditions is established, which reduces all of the plates
simultaneously. It imposes the condition of best fit with respect
to the reference catalogue, but also internal agreement of
overlapping measures. Linear plate solutions in combination with a
corrective mask common to all plates are derived and applied in an
iterative process. The preliminary solutions for each individual
plate are determined using a subset of reference stars identified
to initialize the process.

For the purposes of these CdC-SF plate measures, ``overlapping''
images are broadened to include multiple scans of the same plate,
as well as star images that do actually fall in the overlap area
of adjacent plates. That is, after transforming both A and B scans
onto a common system, and after having applied the various
corrections described in the previous section, the $x$ and $y$
values from each rotated scan will be treated as if they were from
separate plates.

The Tycho-2 Catalogue (H{\o}g et al. 2000) is used as the
reference catalogue, with its proper motions being used to
back-date the coordinates to the epoch of the CdC-SF plates. The
Tycho-2 proper motions have a precision of 2.5 mas/yr and, with a
magnitude limit of V$\sim$11.5, the catalogue provides a
sufficient star density. Abad (1993) has demonstrated that the
overlapping technique does not require a particularly dense grid
of reference stars.

Alternatively, the deeper UCAC2 catalogue could have been used as
a reference, but we have decided against this. While the estimated
precision of the UCAC2 proper motions is only slightly inferior to
that of Tycho-2, it is suspected that its systematic errors may be
significantly larger.  This is due to problems in the ``yellow
sky'' catalogue used as first-epoch material in deriving the UCAC2
proper motions.  The yellow-sky catalogue is based on photographic
plates from the Lick NPM and Yale/San Juan SPM programs, which are
known to suffer from significant but correctible magnitude
equation.  As a sacrifice to expediency, the standard
magnitude-equation corrections adopted by these programs was not
applied during construction of the yellow-sky catalogue.  It is
the bright end of the NPM/SPM plate material that is most
susceptible to these magnitude-equation problems, precisely the
magnitude range in common with the CdC-SF plates. For this reason,
and again noting that it has sufficient star density, Tycho-2 is
our choice of reference catalogue for this application\footnote{We
point out that a revised version of the UCAC2, to be called the
UCAC3, is currently under construction by USNO.  The new proper
motions will be based on an improved reconstruction of the
yellow-sky catalogue, one that makes explicit magnitude-equation
corrections.}.

While the overlap method allows for the reduction of the entire
set of plates, it was decided to divide the plates into four
groups by right ascension. This allows us to compare the results
in areas of low and high star density and to adjust relevant
parameters of the reduction process accordingly. Each group spans
two hours in right ascension and all seven declination zones of
the CdC-SF.

    Not all of the measured plates were included in the final
reduction. Triple-exposure plates have poorer quality than
simple-exposures ones, as demonstrated in the previous section.
Nevertheless, they are, in general, helpful for the overlap
solution in determining field distortion and possible systematic
errors at faint magnitudes. Also, these plates can reinforce the
solution of the simple-exposure plates via the overlapping
conditions. However, in order to avoid spoiling the accuracy of
the final catalogue, only the best triple-exposure plates are
included, those with final measuring error less than 7$\mu$m. The
included set of higher-quality plates has an average
single-measurement internal precision of 0$\farcs$2 in each
coordinate. All of the simple-exposure plates were included except
for a handful that upon manual inspection had failed in part
during the image detection process. In summary, we used 170 of the
180 simple-exposure plates and 100 of the 240 triple-exposure
plates. In the reduction technique, triple-exposure plates are
considered as three independent plates. No blended images are
included.

\begin{figure}[h!]
\centering
\includegraphics[width=7.0cm]{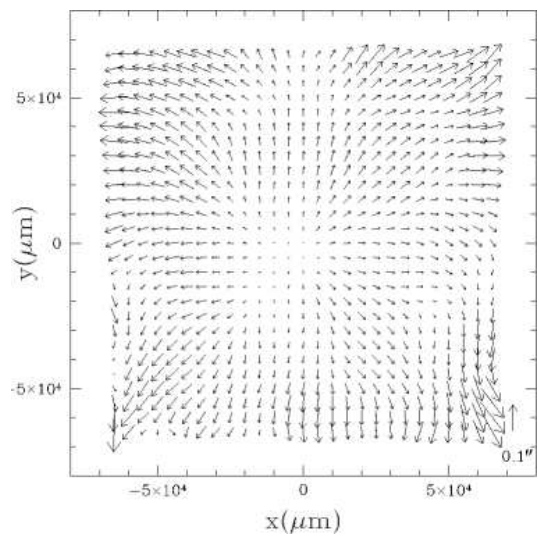}
\caption{Stacked position residuals as a function of coordinates
after only linear plate modelling.} \label{fig15}
\end{figure}

During each iteration, calculated stellar coordinates from the
various plates are averaged to derive a mean position for each
star. Residuals are obtained as differences between individual
positions and their average, if the star is not in the reference
catalogue. If the star is identified as a reference, we also
calculate the residual difference between the average position and
the catalogue position. Both types of residuals from all plates
are then stacked as a function of position and plotted as a vector
field, assigning higher weights to the residuals formed with the
Tycho-2 catalogue than those formed between overlapping plates.
The pattern in Fig. \ref{fig15} shows the stacked residuals, which
are used as a representation of the systematic field distortion
remaining in the plates.  This function is then applied to the
positions and a new iteration of the astrometrical reduction
performed.

\begin{figure*}[ht!]
\centering
\includegraphics[scale=1.25]{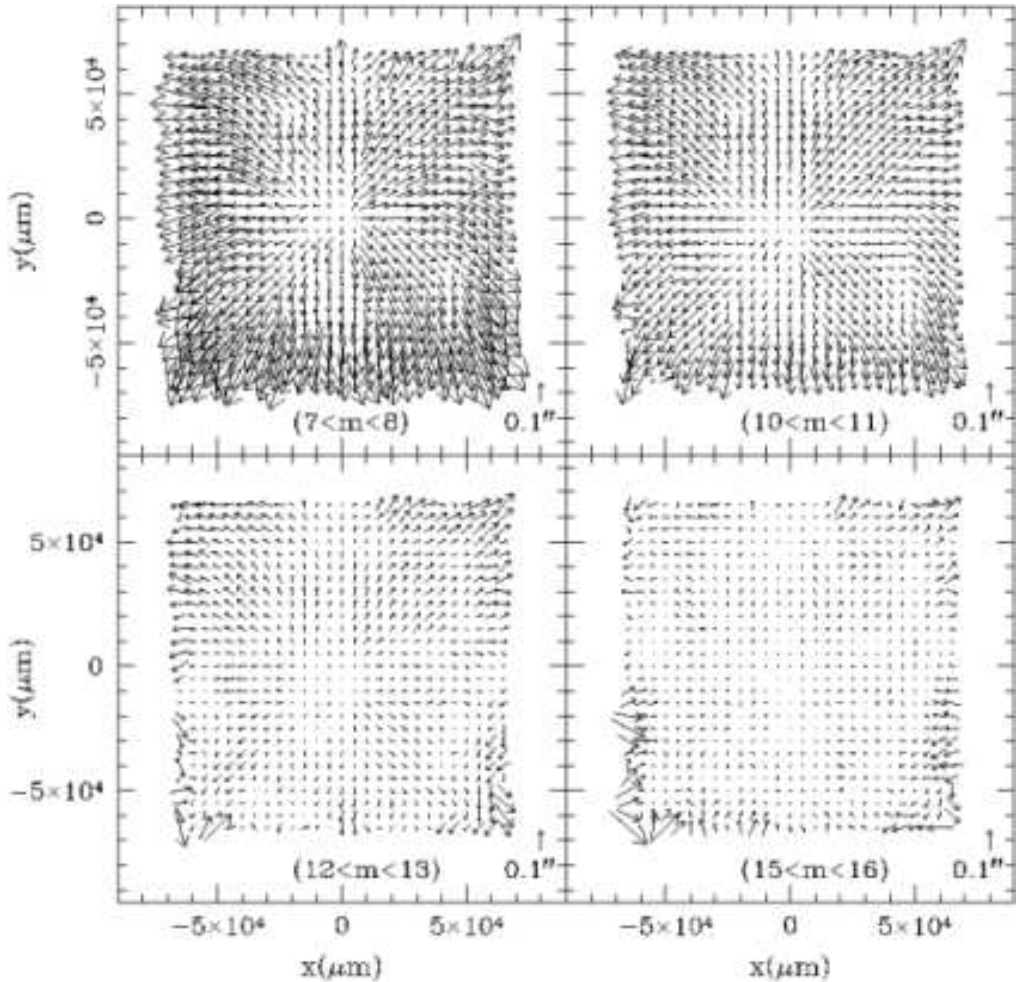}
\caption{Vector residuals are plotted for four different magnitude
ranges. One mask for each magnitude range is derived, and then by
interpolation of these masks, the distortion is corrected for each
position. In this way, systematic error that is a function of
magnitude is removed.} \label{fig16}
\end{figure*}

\begin{figure}[h!]
\centering
\includegraphics[width=7.2cm]{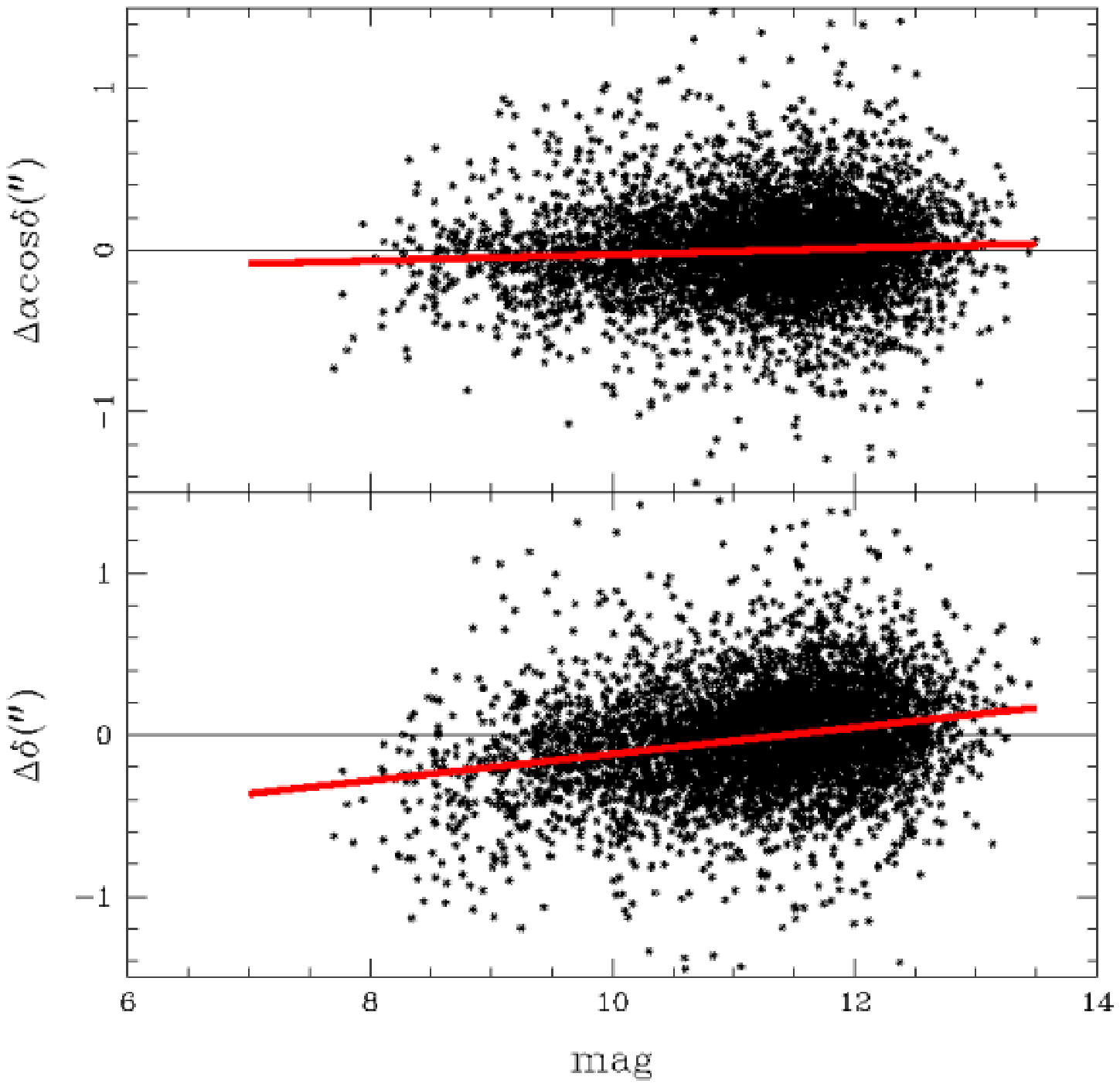}
\caption{Differences in positions for reference stars as a
function of magnitude, where a systematic error is appreciable.
The line is actually a least-squared linear fit to the
differences, indicating that a magnitude equation is still
preset.}
\label{fig17}
\end{figure}

    Different correction masks are derived by binning stars into one-magnitude
wide intervals (Fig. \ref{fig16}) to determine the magnitude
dependence of the systematic errors. Weighted interpolation within
these masks provides the corrections to be applied as a function
of magnitude. It is seen that the distortion is more pronounced at
bright magnitudes. After this correction is made,
a slight residual magnitude equation is
found, fitted, and corrected for, based on differences with the
reference catalogue (Fig. \ref{fig17}).
While the Tycho-2 reference stars do not span the entire magnitude range of
our sample, we feel there is sufficient magnitude overlap, and
thus leverage, to linearly extrapolate the remaining couple of
magnitudes to the faintest stars in our sample.

\section{Evaluation of the final astrometry}

\begin{figure}[h!]
\centering
\includegraphics[angle=90,width=8cm]{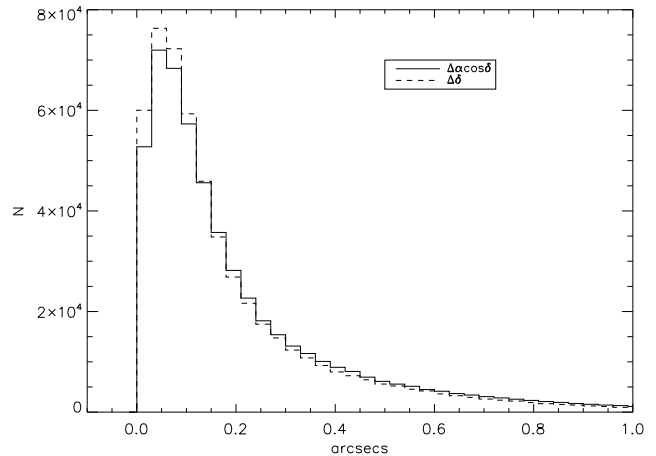}
\caption{Histogram of the internal uncertainties, estimated from the rms of the
positional differences of overlapping images,
for all of the derived CdC-SF star positions. }
\label{fig18}
\end{figure}

An estimated error for each star in our demonstration area is
derived based on the rms of the positional differences of the
images that contributed to the average position. The distribution
of the errors as a function of magnitude for the entire catalogue
is shown in Fig. \ref{fig18} and given in Table \ref{table:2}
separated by right ascension grouping. The mean values of these
uncertainties are $(\sigma_{\alpha cos
\delta},\sigma_{\delta})=(0\farcs21,0\farcs19)$ for the entire
sample, and for stars brighter than 14, the mean values are
$(\sigma_{\alpha cos
\delta},\sigma_{\delta})=(0\farcs12,0\farcs11)$. The final line in
the table also lists the rms differences between our derived
positions and the Tycho-2 catalogue, at the CdC-SF epoch. These
rms differences are expected to have a significant, possibly
dominant, contribution from the Tycho-2 positions at this epoch.

\begin{table*}[ht!]
 \caption{Final internal uncertainty estimates as a function of
magnitude for the derived CdC-SF coordinates, grouped by right ascension.
Also listed are the standard deviations of differences with Tycho-2 positions
at the epoch of the plates. Where $^*$ stands for $cos\delta$.}
 \label{table:2}
 \centering
 \begin{tabular}{|r|rrr|rrr|rrr|rrr|}

 \noalign{\smallskip}
 \hline
 \noalign{\smallskip}

 \multicolumn{1}{|r|}{Area} & \multicolumn{3}{c|}{$06^h\leq\alpha<08^h$} & \multicolumn{3}{c|}{$08^h\leq\alpha<10^h$} &
 \multicolumn{3}{c|}{$10^h\leq\alpha<12^h$} & \multicolumn{3}{c|}{$12^h\leq\alpha<14^h$} \\

 \noalign{\smallskip}
 \hline
 \noalign{\smallskip}
 Magnitude & $\sigma_{\alpha^*} ('')$ & $\sigma_\delta ('')$ & $N_{*'s}$ &
             $\sigma_{\alpha^*} ('')$ & $\sigma_\delta ('')$ & $N_{*'s}$ &
             $\sigma_{\alpha^*} ('')$ & $\sigma_\delta ('')$ & $N_{*'s}$ &
             $\sigma_{\alpha^*} ('')$ & $\sigma_\delta ('')$ & $N_{*'s}$ \\

 \noalign{\smallskip}
 \hline
 \noalign{\smallskip}

    8 &  0.10 &  0.11 &      146 &  0.12 &  0.10 &      114 &  0.12 &  0.16 &       43 &  0.08 &  0.09 &       65 \\
    9 &  0.11 &  0.11 &     1066 &  0.11 &  0.10 &      710 &  0.11 &  0.12 &      369 &  0.11 &  0.10 &      361 \\
   10 &  0.12 &  0.12 &     5437 &  0.12 &  0.11 &     3009 &  0.13 &  0.12 &     1616 &  0.13 &  0.11 &     1368 \\
   11 &  0.11 &  0.10 &    16786 &  0.11 &  0.10 &     7414 &  0.13 &  0.12 &     3445 &  0.13 &  0.11 &     3193 \\
   12 &  0.10 &  0.10 &    34489 &  0.11 &  0.10 &    13918 &  0.13 &  0.12 &     5691 &  0.13 &  0.13 &     5077 \\
   13 &  0.11 &  0.11 &    61426 &  0.12 &  0.11 &    24092 &  0.15 &  0.14 &     8714 &  0.15 &  0.14 &     7843 \\
   14 &  0.22 &  0.20 &   119957 &  0.23 &  0.21 &    46563 &  0.24 &  0.22 &    15161 &  0.26 &  0.24 &    15267 \\
   15 &  0.32 &  0.31 &    84942 &  0.32 &  0.30 &    30315 &  0.31 &  0.30 &     8916 &  0.33 &  0.31 &     9042 \\

 \noalign{\smallskip}
 \hline
 \noalign{\smallskip}

 Total &  0.21 &  0.19 &   324335 &  0.20 &  0.19 &   126142 &  0.21 &  0.19 &    43960 &  0.22 &  0.21 &    42227 \\

 \noalign{\smallskip}
 \hline
 \noalign{\smallskip}

 $\Delta_{Tycho-2}$  &  0.30 &  0.33 & 19066 &  0.34 &  0.35 & 9127 &  0.42 &  0.34 & 4168 &  0.37 &  0.33 & 4028 \\

 \noalign{\smallskip}
 \hline
 \noalign{\smallskip}
 \end{tabular}
\end{table*}

We note that the global plate-overlap solutions involved the use
of both simple and triple-exposure plates to determine the plate
distortions and alignments better. However, in compiling the final
star positions one has the choice of whether to include images
from triple-exposure plates in the position averages. Doing so
will improve completeness while slightly eroding the overall
precision. The uncertainties quoted above were based on inclusion
of the triple images in the final positions. The mean values of
the uncertainties using only the simple-exposure plates for the
final compilation are $(\sigma_{\alpha cos
\delta},\sigma_{\delta})=(0\farcs14,0\farcs14)$ for the entire
sample.

A comparison of our catalogue with Tycho-2 positions at the CdC plates' epoch
as a function of magnitude shows no systematic
pattern, mean differences remaining constant over all magnitudes.
Thus our sample can be considered to have been successfully placed
on the system ICRS, as defined by Hipparcos, via Tycho-2.

Our uncertainty values agree well with results from other groups
deriving astrometry from similar plate material, although from
other CdC collections. Table \ref{table:3} lists the single-plate
precision, as well as the accuracy over extended multi-plate areas
that is obtained in these various studies. The significant
difference between the present study and the others is their use
of much more sophisticated measuring machines to measure the
plates. Specifically, those studies make use of the APM (Rappaport
et al.~2006), a PDS (Ortiz-Gil et al.~1998), and the MAMA (Geffert
et al.~1996). By using an inexpensive flatbed scanner and the
procedures presented here, comparable astrometric precision and
accuracy are obtained. This suggests that limitations intrinsic to
the CdC plate material are determining the final astrometric
precision in all cases.

\begin{table}[h!]
\caption{Comparison of various astrometric studies involving CdC plate material
that employ different measuring machines and reduction procedures.}

\label{table:3}
\centering
\begin{tabular}{llccl}
 \noalign{\smallskip}
 \hline
 \noalign{\smallskip}
Reference         &  Machine  &  Precision     &  Accuracy   & $N_{pl}$ \\
 \noalign{\smallskip}
 \hline
 \noalign{\smallskip}

This paper        &  Scanner  &  $0\farcs18$   & $0\farcs20$ & 400 \\
Rappaport 2006    &  APM      &  $0\farcs15$   & $0\farcs20$ & 512 \\
Ortiz-Gil 1998    &  PDS      &  $0\farcs15$   & $0\farcs15$ & 1 \\
Geffert   1996    &  MAMA     &  $0\farcs15$   & $0\farcs20$ & 2 \\
 \noalign{\smallskip}
 \hline
 \noalign{\smallskip}
\end{tabular}
\end{table}

Another property of our demonstration sample that needs to be
characterised is its completeness. This will obviously be an
important aspect of any future catalogue to be based on scanner
measures of CdC plates. Our demonstration catalogue contains
positions and estimated uncertainties for approximately 560\,000
stars that have been selected using the following criteria: (1)
the star must match with a counterpart in the UCAC2 catalogue
within 3$\farcs$5 of tolerance at the CdC plate epoch, (2)
duplicate measures, from different plates, within 3$\farcs$0 of
tolerance are combined into a single entry by averaging the
positions, and (3) the star must appear in at least 2 different
plate scans, ensuring the minimum constraint that each star's
image be present in both scan rotations.

\begin{figure}[h!]
\centering
\includegraphics[width=8cm,height=6cm]{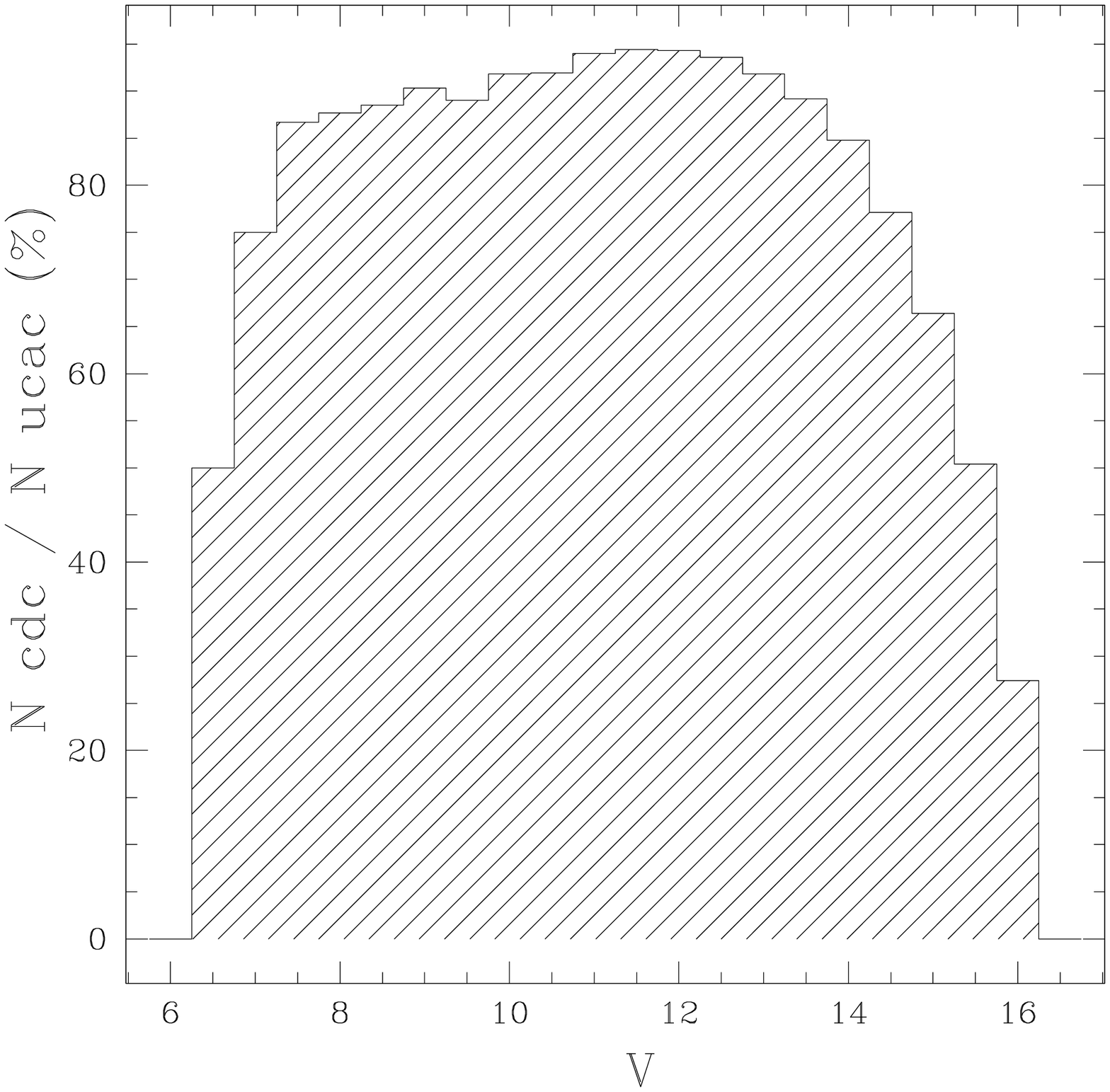}
\includegraphics[width=8cm,height=5cm]{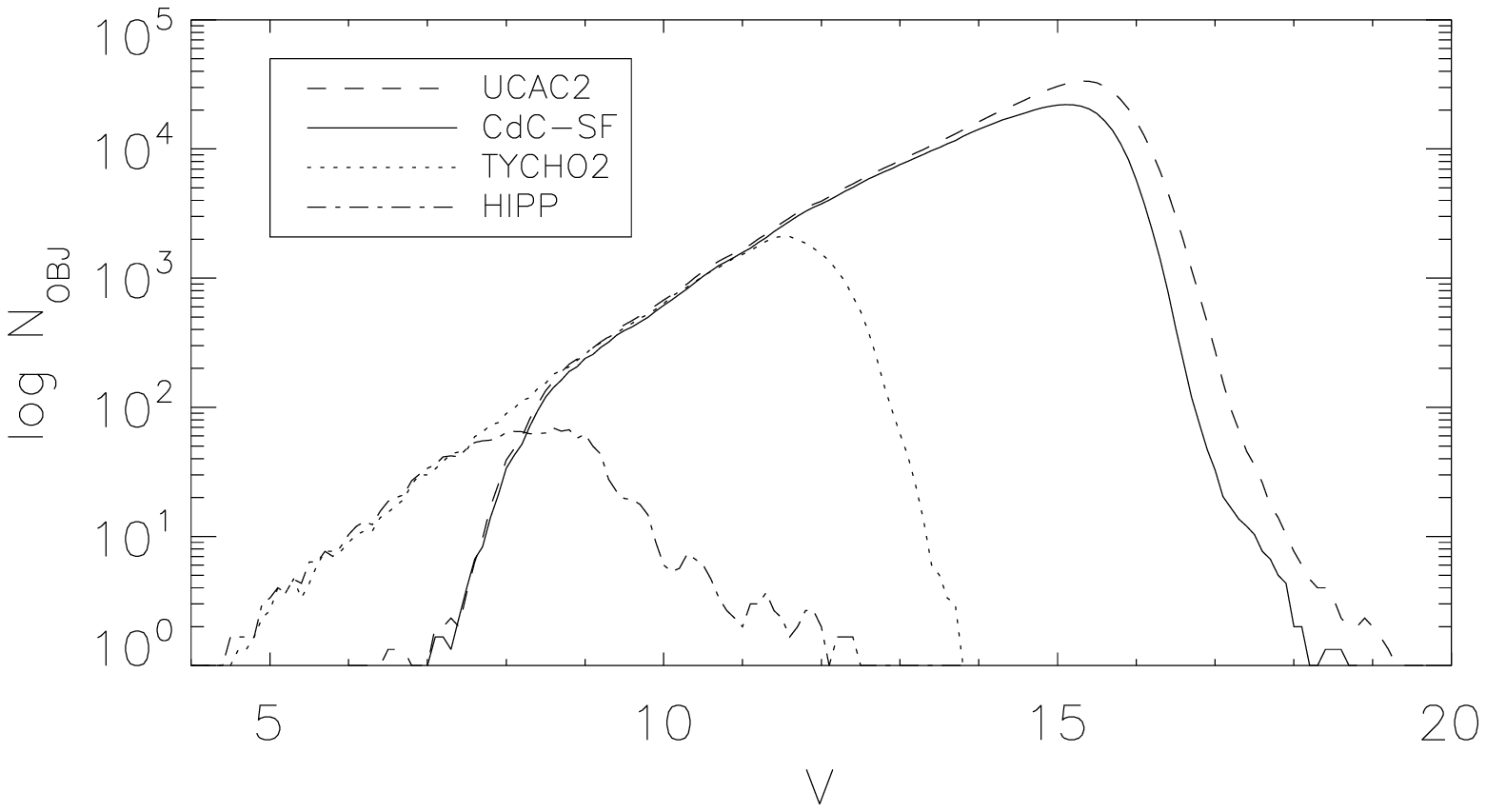}
\caption{ Completeness of our CdC-SF demonstration area sample.
{\it Top -} The percentage of stars in the sample compared to the
input list from the UCAC2 Catalogue is given as a function of
magnitude. {\it Bottom -} The magnitude distribution of our CdC-SF
sample is compared to other astrometric catalogues; Hipparcos,
Tycho-2, and UCAC2. } \label{fig19}
\end{figure}

The input catalogue from which the list of CdC-SF objects has been
derived is based on the UCAC2 catalogue. For this reason the upper
limit of our area's completeness is set by that of the UCAC2.
However, the properties and condition of the plates and the number
of plates not included in the reduction, as well as the automated
nature of the reduction pipeline, produce additional losses.
Figure \ref{fig19}a shows the relative completeness of CdC-SF
compared to UCAC2. Presumably, modifications to the automated
pipeline could improve the completeness of our procedures, most
notably on the bright end. For instance, bright stars that failed
to centre because of extreme saturation may be recovered by
fitting with the tepui function (Vicente \& Abad 1999)
specifically developed for such saturated profiles.

The magnitudes used in constructing both panels of Fig.
\ref{fig19} are extrapolated estimates of $V$ made from 2MASS
$J,K$ photometry that is included in the UCAC2 catalogue. This
so-called pseudo-$V$ follows the empirical relation given by
Girard et al.~(2004)

\begin{center}
$V_{pseudo} \approx J + 2.79\ (J-K).$
\end{center}

\noindent This approximation works reasonably well over a range of
spectral types and allows us to compare the magnitude
distributions of various catalogues on a common system, such as is
shown in Fig. \ref{fig19}b. Our CdC-SF sample is compared to the
Hipparcos, Tycho-2, and UCAC2 catalogues extracted over the same
area of sky. The faint-end turnover in the (logarithmic)
distribution of our CdC-SF sample is at $\sim$V=15.1.

As an aside, we note that a crude magnitude estimate was
calculated for all stars in our sample, based on calibrating the
instrumental magnitude estimates of the Gaussian centering
algorithm with the admittedly very approximate $R$ magnitudes of
the UCAC2 catalogue. A simple polynomial transformation was
adopted, and this gave poor results due, presumably, to both the
mismatch in passbands and the crude UCAC2 photometry used as
reference. These magnitudes were nonetheless adequate as
indicators of magnitude when required in the astrometry reduction
pipeline. They are also the magnitudes used in all plots, other
than Fig. \ref{fig19}, presented in this paper, whenever residuals
are shown as a function of magnitude.  For this they are also
adequate.

Our primary emphasis is astrometry, with the goal of eventually
producing an absolute proper-motion catalogue.  Photometry is of
secondary priority. In order to eventually provide reasonable
photographic photometry, a program is currently underway to
calibrate the scanner-based instrumental magnitudes using
observations made with the 0.8-m Telescope IAC-80 (Observatorio
del Teide, Tenerife). At this time, we do not have enough
observations (just five fields of $11\farcm3\times11\farcm3$) to
cover our demonstration area. In the future, a combination of
these standards, along with brighter, secondary calibrations,
e.g., Tycho-2 stars, will provide a means of calibrating our
scanner-based instrumental magnitudes better.

\section{Conclusions and future plans}

The 1260 original {\it Carte du Ciel} plates, San Fernando zone,
have been resurrected in a digital form by use of a commercial
flatbed scanner. A method of removing the mechanical distortion
introduced by the scanner is presented. A final measuring accuracy
of 0$\farcs$2 is achieved, similar to what has been obtained in
other studies using specialized plate-measuring machines for
similar plate material.

Currently, one third of the CdC-SF collection has been measured
and reduced (420 plates). This demonstration area has yielded
positions at epoch $\sim$1901.4 for $\sim$560\,000 stars, covering
a total area of 1080 deg$^2$ in the sky. This sample has a
completeness of 85\% in the range 7.0 $\leq$ V $\leq$ 14.5. The
positions are on the ICRS system defined by Tycho-2 at the epoch
of the observation. Internal error estimates are based on multiple
exposures for the same star from overlapping plates (0$\farcs$2).
External comparison with the Tycho-2 positions (rms of the
differences of 0$\farcs$3) shows that the quality of our measures
is as good or better than obtained for the AC project, but extends
to fainter magnitudes.

Thus, internal precision and external uncertainty estimates that match
those attained with more precise machines for CdC plates
demonstrate the potential of this flatbed-scanning and reduction method
for exploiting these historical plates.

The early epoch positions derived in our demonstration area are
currently being combined with UCAC2 modern positions to calculate
absolute proper motions that make use of this long-time baseline.
Additionally, the full CdC-SF zone, all 24 hours of right
ascension, has been digitized with the flatbed scanner described
here and will be reduced following similar procedures. Ultimately,
the resulting astrometric catalogue will provide a useful tool in
the examination of Galactic structure and kinematics.

\begin{acknowledgements}
We are very grateful to the Observatorio de San Fernando for
making the Carte du Ciel plates available to us from their
historical archive. We also want to thank all of the people who
have participated in the digitization of the collection, with
special mention to Jos\'e Mui\~nos, Fernando Beliz\'on and Miguel
Vallejo.

The authors wish to thank Terry Girard of Yale University (USA)
for providing us with portions of the software used in this work and
for very useful discussions and comments during this study.
\end{acknowledgements}


\begin{thebibliography}{}

\bibitem[Abad(1993)]{} Abad, C.\ 1993, \aaps, 98, 1

\bibitem[Abad \& Vicente(1999)]{} Abad, C., \& Vicente, B.\ 1999, \aaps, 136, 307

\bibitem[Auer \& van Altena(1978)]{} Auer, L.~H., \& van Altena, W.~F.\ 1978, \aj, 83, 531

\bibitem[Bertin \& Arnouts(1996)]{} Bertin, E., \& Arnouts, S.\ 1996, \aaps, 117, 393

\bibitem[Dick et al.(1993)]{} Dick, W.~R., Tucholke, H.~J., Brosche, P., Galas, R., Geffert, M.,
\& Guibert, J.\ 1993, \aap, 279, 267

\bibitem[Geffert et al.(1996)]{} Geffert, M., Bonnefond, P., Maintz, G., \& Guibert, J.\ 1996, \aaps, 118, 277

\bibitem[Gill(1898)]{} Gill, D.\ 1898, \mnras, 59, 61

\bibitem[Girard et al.(2004)]{} Girard, T.~M., Dinescu, D.~I., van Altena, W.~F., Platais, I.,
Monet, D.~G., \& L{\'o}pez, C.~E.\ 2004, \aj, 127, 3060

\bibitem[H{\o}g et al.(2000)]{} H{\o}g, E., Fabricius, C., Makarov, V.,
Urban, S., Corbin, T., Wycoff, G., Bastian, U., Schwekendiek, P. \& Wicenec, A.\ 2000, \aap, 357, 367

\bibitem[Lamareille et al.(2003)]{} Lamareille, F., Thi{\'e}vin, J., Fournis, B., Grimault, P., Broquet, L.,
\& Davoust, E.\ 2003, \aap, 402, 395

\bibitem[Lattanzi et al.(1991)]{} Lattanzi, M.~G., Massone, G., \& Munari, U.\ 1991, \aj, 102, 177

\bibitem[Lee \& van Altena(1983)]{} Lee, J.~F., \& van Altena, W.~F.\ 1983, \aj, 88, 1683

\bibitem[Ortiz-Gil et al.(1998)]{} Ortiz-Gil, A., Hiesgen, M., \& Brosche, P.\ 1998, \aaps, 128, 621

\bibitem[Rapaport et al.(2006)]{} Rapaport, M., Ducourant, C., Le Campion, J. et al.\ 2006, \aap, 449, 435

\bibitem[Stock(1981)]{} Stock, J.\ 1981, RMxAA, 6, 115

\bibitem[Stock \& Abad(1988)]{} Stock, J., \& Abad, C.\ 1988, RMxAA, 16, 63

\bibitem[Urban et al.(1998)]{} Urban, S., Corbin, T., Wycoff, G., Martin, J., Jackson, E.,
Zacharias, M., \& Hall, D.\ 1998, \aj, 115, 1212

\bibitem[Vicente \& Abad(2003)]{} Vicente, B., \& Abad, C.\ 2003, Astronomy in Latin America, 31

\bibitem[Zacharias et al.(2004)]{} Zacharias, N., Urban, S., Zacharias, M., Wycoff, G.,
Hall, D., Monet, D., \& Rafferty, T.\ 2004, \aj, 127, 3043

\end{thebibliography}
\end{document}